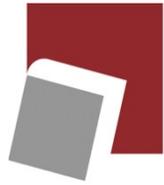

Institute for Advanced Studies in Basic Sciences
Gava Zang, Zanjan, Iran

# Investigating the stellar system's life-time and the evolution of their mass function using N-body simulation

Master's Thesis

**Zeinab Khorrami**

**Supervisor: Dr. Hossein Haghi**

Winter 2011

# Abstract

In this thesis we study several aspects of dynamical evolution of stellar clusters. The results of more than 200 simulations of single-mass star clusters with different initial total mass, half-mass radius and galactocentric distance, are reported. Recent studies of star clusters show a linear relation between a star cluster's dissolution time and its two-body relaxation time in logarithmic scale. We found that the single-mass star clusters do not show such a linear relation. We present new modified initial parameters to obtain a linear relation for single-mass star clusters.

Also the evolution of multi-mass clusters and their lifetime, in the presence of the Galaxy is investigated. We simulate about 90 multi-mass star clusters with the Nbody6 code. These clusters have different initial total mass, half-mass radius and galactocentric distance. Finally we investigate the evolution of the stellar mass function and show that the slopes of the mass functions decrease with time. In addition we study the effect of galactocentric distance of star clusters on the evolution of the mass function.

# Contents

# Chapter 1

# Introduction

**1. Star clusters**

Star clusters are groups of similar stars (in terms of age and chemical composition) which are bound by mutual gravitational attraction. The stars within a star cluster are approximately of the same age and distance and this makes star clusters unique laboratories for studies of stellar evolution. The member stars of the cluster have the same metallicity because they are formed from the same molecular cloud. There are two distinguishable types of star clusters; globular star clusters (GC) and open star clusters (OC).

Globular Clusters are nearly symmetrical systems of up to a million stars formed about 13 to 15 billion years ago, with the highest concentration of stars near its own center. They are the oldest surviving stellar subsystems in galaxies. The number of stars in globular clusters is larger than open clusters. Massive globular clusters exert a strong gravitational attraction on their members and can survive for many billions of years but open clusters become disrupted by close encounter with other clusters and clouds of gas as they move around the galaxy center and they generally survive for a few hundred million years.

Individual clusters as well as GC systems are of great worth as specific targets but also represent a powerful tool to obtain a deep insight into a large variety of astrophysical and cosmological problems (see table 1. 1).[1]

| Subject | Reasons for importance |
|---|---|
| Witnesses of the early Galactic evolution | First to form<br><br>Chemically uncontaminated |
| Stellar Evolution Laboratories | Simple stellar populations<br><br>Test of the 'stellar clock' |
| Distance indicators | Standard candles: the RR Lyrae stars<br><br>GC system integrated luminosity function |
| Age indicators | The turn-off luminosity = 'the clock'<br><br>absolute ages: lower limit to the age of the universe<br><br>relative ages: 'second parameter' and Galaxy formation and evolution |
| Dynamics probes | Dense environment (core collapse / evaporation / collisions / merging–surviving / segregation )<br><br>Test particle of the galactic gravitational field |
| Containers of peculiar objects | X-ray sources (strong–weak–diffuse) / Blue stragglers / Binaries / Planetary nebulae / White dwarfs / Cataclysmic variables / Millisecond pulsars / Neutron stars |

*Table 1. 1*

---

Stars and star clusters are formed from giant molecular clouds (GMCs). These cold (10 to 20 K) clouds of gas are heavy enough (average density of $10^2 - 10^3 \ (particles/cm^3)$ ) to collapse under their own gravity but such a heavy cloud cannot collapse into a single star.

Due to relatively high stellar densities, interesting dynamical events, such as stellar collisions and binary formation and disruption, take place in a cluster.

**2. Dynamical evolution**

Star clusters are the ideal astrophysical objects to study the stellar dynamics and evolution. The current properties of star clusters enable us to learn about their formation and evolution. Recent studies of star clusters solved many problems, however, the complete knowledge of their initial structure and formation still does not exist. The important process which determines the evolution of star clusters is two-body relaxation. In the case of Globular Clusters, its time scale is shorter than cluster ages, and therefore, Globular Clusters are well relaxed spherical systems with quasi-stationary distributions of stellar velocities. However, other processes, like for instance interaction with the tidal field of the galaxy, may be also significant, and they can for example speed up the dissolution of the cluster if it passes through the galactic disk or close to the central bulge[2]. Other relevant physical processes which can affect the evolution of star clusters are encounters with giant molecular clouds and spiral arms (Gieles *et al.* 2006, 2007).

In the cluster, the energy (kinetic and potential) of individual stars changes in encounters with other stars and binary stars. If the energy of some star becomes positive, the star escapes from the cluster. Since the two-body relaxation continuously re-establishes a quasi-stationary distribution of stellar velocities (given by the Maxwellian statistics), there are always stars with positive energies, i.e. velocities larger than the escape velocity from the cluster. The continuous escape of such stars from the cluster is called evaporation. It has been shown that isolated clusters have very long evaporation times. However, it can be decreased if the tidal field of the host galaxy is taken into account. It is because the tidal field decreases the escape velocity from the cluster and makes the evaporation faster.

---

2  Vesperini, E., 2009, *Star cluster dynamics*

Theoretical studies and observations on clusters have shown that "mass loss due to two-body relaxation leads to the preferential escape of low-mass-stars" (Koch *et al.* 2004). It results in flattening of the slope of the stellar mass function. If we assume that clusters are formed with the universal IMF, their current mass function provides us with information about their dynamical evolution history. Observations of a number of globular clusters show that different clusters have different slopes of the mass function. It may be the result of their different dynamical evolution, in particular, different processes of dissolution.

Another piece of evidence for the preferential loss of low-mass stars can be seen in the evolution of the mass-to-light ratio which is lower than the value predicted by models which consider only stellar evolution. It is because low mass stars have negligible luminosity, so, if they escape from the cluster the total mass decreases, however, the luminosity is not affected.

## 3. This thesis

In this thesis more than 300 clusters with different initial conditions were simulated with the Nbody6 code[3]. Although this code is capable of including stellar evolution and similar effects, they were not implemented here to focus on pure dynamical evolution. In chapter 2, the initial condition and the setup

---

3  The simulations of this work were done in the computer center of the *Institute for Advanced Studies in Basic Sciences (IASBS)* in Zanjan. [*4CPUs(2664M Hz), mem : 8071448.000KB*] of the cluster is described, then, some important parameters which are used in the input files are explained. At the end of chapter 2 some outputs of this code with their graphical examples is shown.

In chapter 3 , the results of more than 200 simulations of single-mass star clusters with different initial total mass, half-mass radius and galactocentric distance, are reported. Our main focus is to find the relation between the dissolution time of these clusters and some initial parameters like initial total mass ( $M_0$ ) , initial half-mass radius ( $r_{h0}$ ) and galactocentric distance ( $R_G$ ).

Recent studies of star clusters show a linear relation between a star cluster's dissolution time and its two-body relaxation time in logarithmic scale. Single-mass star clusters do not show such a linear relation. We present new initial parameters to obtain a linear relation for single-mass star clusters.

In chapter 4, the evolution of the mass function for multi-mass clusters and their lifetime, in the presence of the Galaxy is investigated. In this section, more than 90 multi-mass star clusters are simulated with the Nbody6 code. These clusters have different initial total mass, half-mass radius and galactocentric distance. The evolution of the mass function's slope shows that these values decrease. In addition we study the effect of galactocentric distance of star clusters on the evolution of the mass function.

# Chapter 2

# N-body simulation

**1. Introduction**

N-body6 is a code to solve the N body problem. This code can create real size models of star clusters with realistic distributions of stellar masses and including all the relevant physical processes, from cumulative distant encounters to close encounters and collisions, to the formation and disruption of binary systems, the effect of external tidal fields (static or varying) and the effect of stellar evolution. These sophisticated models, in turn, demand detailed observations of the structure and stellar content of rich clusters, which can both guide and constrain theories of their formation and evolution.

In addition to their role as stellar dynamics laboratories, star clusters provide clues to reconstruct the early stages of the evolution of our Galaxy and ultimately of other galaxies as well.

This aim, requires a good knowledge of the dynamical evolution of clusters both from a theoretical and an observational point of view.

## 2. Setup of the cluster

It is assumed that the cluster moves on a circular orbit in the Galaxy, with a rotational speed corresponding to distance from the galaxy center. (The cluster which is in a distance about 8.5 Kpc from galaxy center, has a rotational speed about 220 km/s) .

The computations are conducted in a rotating reference frame, which moves with the cluster's initial angular velocity, with the origin of the coordinate system lying in the initial barycenter of the cluster. The x-axis points to the Galactic center and the y-axis is parallel to rotational velocity, while the z-axis is perpendicular to the orbit of the cluster. (Figure 2. 1)

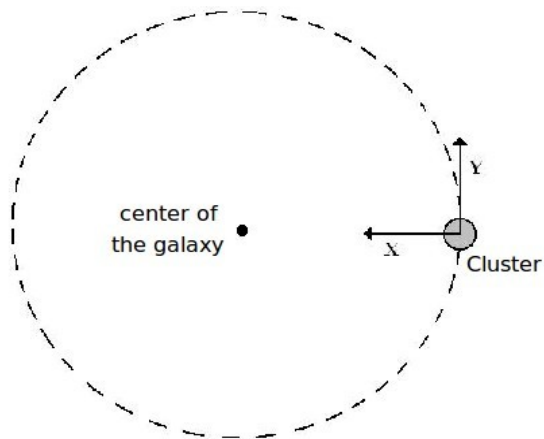

*Figure 2. 1*

We simulate both single-mass and multi-mass star clusters. In single-mass clusters, all the stars have the same mass equal to one solar-mass but in multi-mass clusters, stellar masses are set by a distribution function within a given range.

Two density profiles (Plummer and King model) are used to derive the initial position of stars in the cluster. Initial velocity distribution of the stars in the cluster is supposed to be Maxwellian. It means that there is a high-velocity tail of stars so each star which has a velocity greater than the escape velocity can leave the cluster.

## 3. The N-body6 input file

In this section we describe parameters stated in the input file which are relevant to this work. All values have to be given in a following system of units. Masses are measured in Solar-Masses, distances are measured in parsecs (pc) and time is measured in millions of years (Myr).

Gravitational constant is:

$$G = (6.67259 \mp 0.00085) \times 10^{-11} \left( \frac{m^3}{kgs^2} \right) \quad (1)$$

$$\begin{aligned} m &= 3.24078 \times 10^{-17} \, pc \\ kg &= 5.02785 \times 10^{-31} \, M_\odot \\ s &= 3.16888 \times 10^{-14} \, Myr \end{aligned} \quad (2)$$

From equation 1 and 2 , the gravitational constant is equal to :

$$G = (4.49831 \mp 0.00057) \times 10^{-3} \left( \frac{pc^3}{M_\odot \, Myr^2} \right) \quad (3)$$

Also from equation 2 it can be seen that the velocities can be measured in both km/s and pc/Myr because they are roughly equal.

$$\left( \frac{km}{s} \right) = 1.02269 \left( \frac{pc}{Myr} \right) \quad (4)$$

### 3. 1. Mass function of stars

The star cluster can be chosen as a single-mass or multi-mass cluster. In the single-mass star clusters, mass function is a delta function which is:

$$n \sim \begin{cases} m & , \, m = M_\odot \\ 0 & , \, m \neq M_\odot \end{cases} \quad (5)$$

The mass of every star is equal to one solar-mass. So the total mass of the cluster is equal to total number of the stars in the cluster.

In multi-mass star clusters, every star cluster has a mass function. This mass function describes the mass distribution (The histogram of stellar masses). The initial mass function (IMF) can be chosen as Kroupa initial mass function which is:

$$N\,dm \sim \begin{cases} m^{-1.3}\,dm &, m_{min}<0.5 \\ m^{-2.3}\,dm &, 0.5<m_{max} \end{cases} \quad (6)$$

where N(m) dm, is the number of stars with masses in the range m to m + dm .

### 3. 2.    Initial total mass of the star cluster

In this part the initial total mass of the cluster is given in solar masses.

### 3. 3.    Initial density profile of the cluster

The density profile is used to describe the distribution of positions and velocities of particles in dynamical systems. Both Plummer spheres and King profiles can be chosen as a initial density profile. In the Plummer sphere

$$\rho(r)=\frac{3M_0}{4\pi a^3}\left(1+\frac{r^2}{a^2}\right)^{-5/2} \quad (7)$$

In this equation r is the distance from star to cluster center, a is the core radius, which defines the plummer sphere and $M_0$ is the total mass of the cluster. Integrating this formula gives us the mass distribution:

$$M(r)=\frac{M_0 r^3}{(r^2+a^2)^{3/2}} \quad (8)$$

### 3. 4.    Initial stellar velocities

From the mass distribution, we can derive the potential :

$$\Phi(r) = -\frac{GM}{a}\left(1 + \frac{r^2}{a^2}\right)^{-1/2} \quad (9)$$

The velocity dispersion is calculated by manipulating the Jeans equations [1]:

$$\sigma^2 = \frac{1}{\rho}\int_r^\infty \rho(r)\,\frac{d\Phi_{total}}{dr}\,dr = -\frac{1}{6}\Phi(r) \quad (10)$$

### 3. 5. Initial half-mass radius of the star cluster

Half-mass radius is the radius from the core that contains half the total mass of the cluster. In our simulations the initial half-mass radius is varied from 0.5 pc to 6 pc.

### 3. 6. Tidal field

The tidal radius is a distance from the cluster center at which the gravitational attractive force of the cluster acting on a star is overbalanced by the gravitational force of the galaxy. Every star in the clusters feels two gravitational forces, one from the cluster and one from the galaxy (which the cluster moves round). When stars go further from the center of the cluster, the galaxy's gravitational force becomes stronger or more noticeable than the cluster's gravitational force, so stars can leave the cluster (or when they already have the necessary energy for escape (Baumgardt 2001) ). The radius where the galactic gravitational force become larger than the cluster's gravitational force, is the tidal radius. Inside the tidal radius stars belong to the clusters but outside this radius stars can leave the cluster.

Nbody6 code suggests 3 tidal forces:

1. near-field approximation

2. point-mass galaxy

In this section, it is assumed that the total mass of the galaxy (about one hundred billion solar-

---

[1] Binney, J, & Tremaine, S., "*Galactic Dynamics*", Princeton Series in Astrophysics

masses) concentrate in a point in the center of the galaxy.

3. Allen-Santillan tidal field

The Allen & Santillan (1991) MW potential is completely analytical. This model consists of a spherical central bulge and a disk, both of the Miyamoto-Nagai (1975) form, plus a massive, spherical halo. The total mass of the model is $9 \times 10^{11}$ solar masses. The model escape velocity for objects in the solar vicinity is 535.7 km/s .

For central bulge:

$$V_1(w,z) = -\left(\frac{M_1}{\left(w^2+z^2+b_1^2\right)^{1/2}}\right) \quad (11)$$

$$M_1 = 606.0, b_1 = 0.3873$$

For disk:

$$V_2(w,z) = -\left(\frac{M_2}{\left(w^2+\left(a_2+\left(z^2+b_2^2\right)^{(1/2)}\right)^2\right)^{1/2}}\right) \quad (12)$$

$$M_2 = 3690.0, a_2 = 5.3178, b_2 = 0.2500$$

And for halo:

$$V_3(R) = -\left(\frac{M(R)}{R}\right) - \left(\frac{M_3}{1.02\, a_3}\right)\left[\left(\frac{-1.02}{1+(R/a_3)^{1.02}}\right) + \left(\ln\left(1+(R/a_3)^{1.02}\right)\right)\right]_R^{100} \quad (13)$$

$$M(R) = \frac{M_3 (R/a_3)^{2.02}}{1+(R/a_3)^{1.02}}$$

$$M_3 = 4615.0, a_3 = 12.0$$

In our simulations we use Allen-Santillan tidal field which is a representation of the Milky-Way

Galaxy, so it is easy for comparing our results with observed clusters in our Galaxy.

### 3. 7. Initial position and velocity of the star cluster

In this section the initial position of the cluster is given, in Nbody6's frame. For example if a cluster is at a galactocentric distance about 8.5 kpc in the disk of the galaxy, the initial position will be:

$$\vec{R} = (8500, 0, 0)$$

Also the magnitude and direction of the initial velocity of the cluster is characterized. For example if the initial velocity of a cluster is 220 km/s in y-direction, it will be:

$$\vec{V} = (0, 220, 0)$$

Finally, we list values of the above parameters used in this work. Table 2. 1 shows values used in simulations of single-mass clusters (described in Chapter 3), table 2. 2 shows the same for multi-mass clusters (Chapter 4).

| Input Parameters | Initial value |
|---|---|
| Mass function of the stars | Delta function (eq. 5) |
| Total mass of the cluster | 1000 to 7000 Solar-mass |
| Density profile of the cluster | Plummer model (eq. 7) |
| Half-mass radius | 0.5 to 6 pc |
| Tidal field of the galaxy | Allen-Santillan (eq. 11 to 13) |
| Initial position of the cluster | $\vec{R} = (8500, 0, 0)$ To $\vec{R} = (40000, 0, 0)$ |
| Initial velocity of the cluster | $\vec{V} = (0, 220, 0)$ |

*Table 2. 1 : Initial values for single-mass clusters (chapter 3)*

| **Input Parameters** | **Initial value** |
| --- | --- |
| Mass function of the stars | Kroupa IMF (eq. 6) |
| Total mass of the cluster | 3000 to 10000 Solar-mass |
| Density profile of the cluster | Plummer model (eq. 7) |
| Half-mass radius | 0.5 to 6 pc |
| Tidal field of the galaxy | Allen-Santillan (eq. 11 to 13) |
| Initial position of the cluster | $\vec{R} = (8500, 0, 0)$ To $\vec{R} = (35000, 0, 0)$ |
| Initial velocity of the cluster | $\vec{V} = (0, 220, 0)$ |

*Table 2. 2 : Initial values for multi-mass clusters (chapter 4)*

**2. Output file summery for N-body6 and some examples**

After simulating star clusters with initial conditions and mentioned inputs, outputs which contain the situation of the cluster in different time-steps, are saved in output files. There are three types of output files. The first one includes the overall cluster properties like the cluster mass, different lagrange radii, total energy and number of binary stars etc. The second one includes all the information about individual stars (in particular their positions, velocities and masses) in given times. The third one includes the information about the escaper stars. Some important outputs are mentioned here:

1. **mass and number of stars**

    In the first output file, time is the first column which starts from time 0 and describes the cluster at 10Myr time-steps. The last line of this file is related to the time at which the cluster's life is over and the cluster has nearly lost all its stars. The second column gives the cluster's total mass and the third column declares the number of the cluster's stars. For example if a single-mass cluster with initial mass about 1000 solar-mass is simulated, in the output file, in the first line time(first column) is 0, mass (second column) is 1000 and number of stars (third column) is 1000. Since the cluster loses its mass and stars as time passes, if mass (column 2) is plotted

versus time (column 1), the evolution of the cluster can be seen easily.

In figure 2. 2, the evolution of three single-mass clusters with initial half-mass radius about 3 pc is shown. These clusters are at a galactocentric distance of about 8.5 kpc in the Allen-Santillan tidal field. The key in the upper right hand corner shows which color represents which cluster. For example the green line shows evolution of the single-mass cluster with initial mass of 5000 solar-masses and the blue line is for the single-mass cluster with initial mass of 3000 solar-masses.

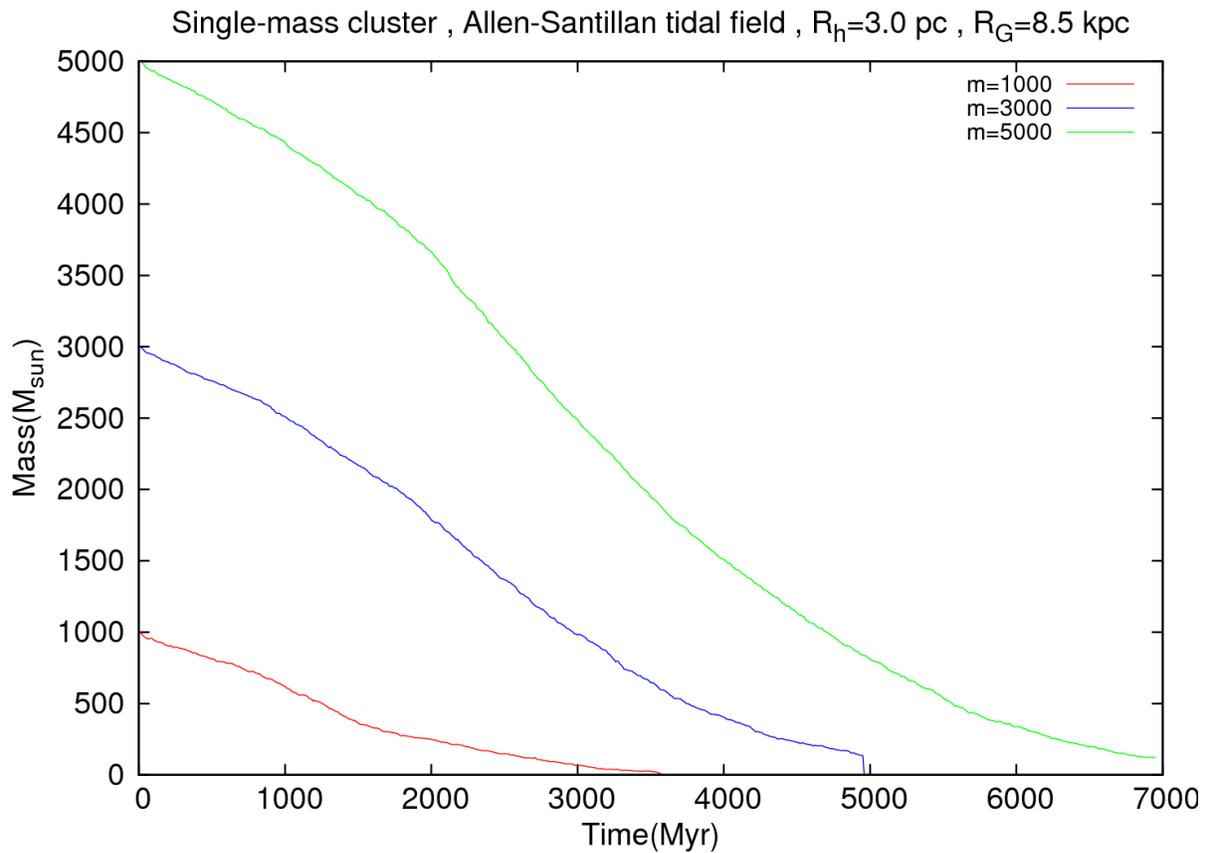

*Figure 2. 2*

2. **Inner and outer radii**

    Lagrange radii which include 2, 5, 10, 20 percent of the total mass of the cluster, are called

inner radii. Columns 19 to 22 in the output file describe the evolution of these inner radii. Column 24 is for the Lagrange radius which includes 90 percent of total mass of the cluster which is considered as a outer radius.

In the beginning the cluster is very compact and sits deeply within the tidal sphere. As the core immediately starts collapsing, the outer layers expand and the inner layers contract. Core-collapse is reached at the time where the cluster has expand sufficiently to fill its tidal radius and starts spilling over the tidal radius. By plotting different radii, the core-collapse time is detectable. In figure 2. 3, the evolution of different radii is plotted.

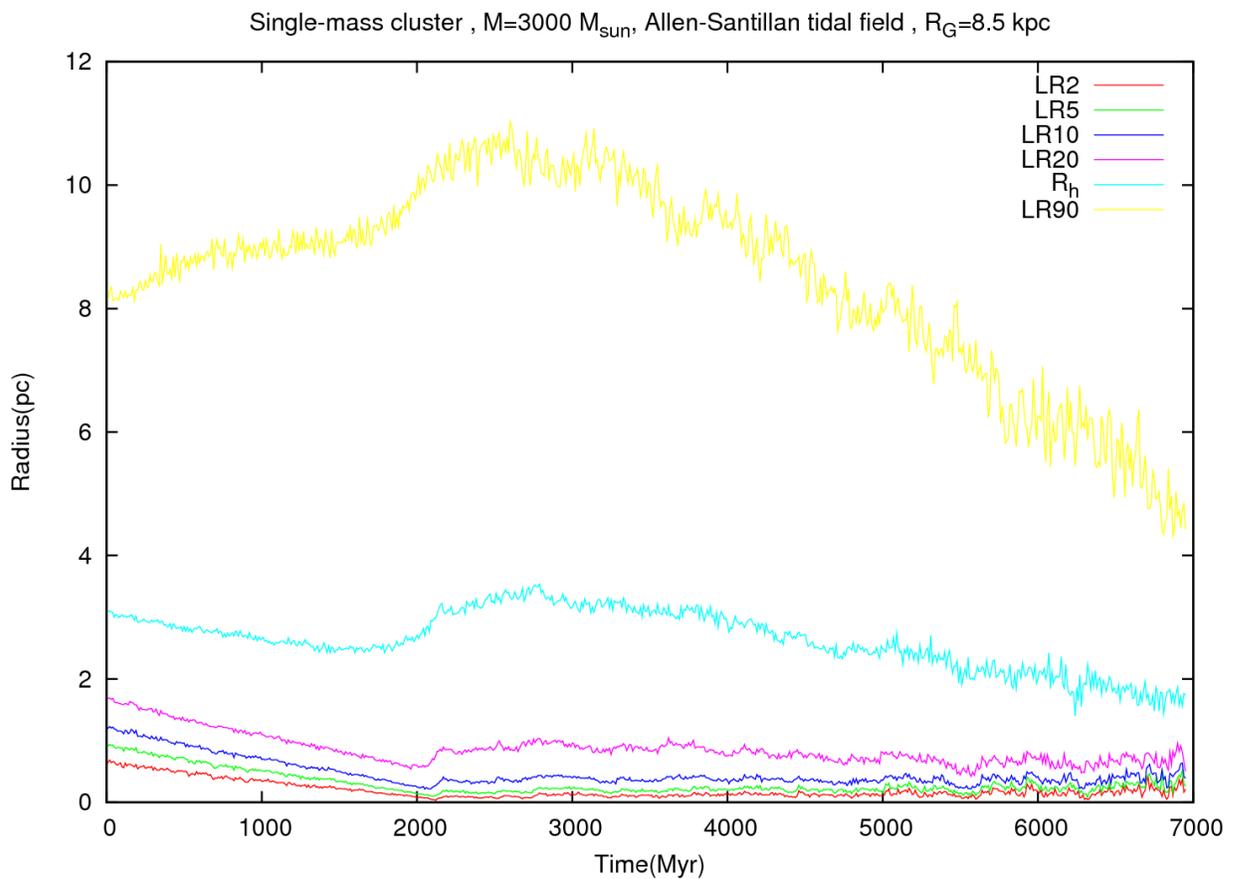

*Figure 2. 3*

This single-mass cluster has a initial total mass of 3000 solar-masses and galactocentric distance of 8.5 kpc in the Allen-Santillan tidal field. The key in the upper right hand corner shows which color is for which Lagrange radius.

**3. Half-mass radius**

One of the important values in the study of the evolution of star clusters is the half-mass radius. The evolution of the half-mass radius can be studied by plotting column 23 (half-mass radius) versus time. In figure 2. 4 the evolution of half-mass radius for similar single-mass clusters with different initial half-mass radii is plotted.

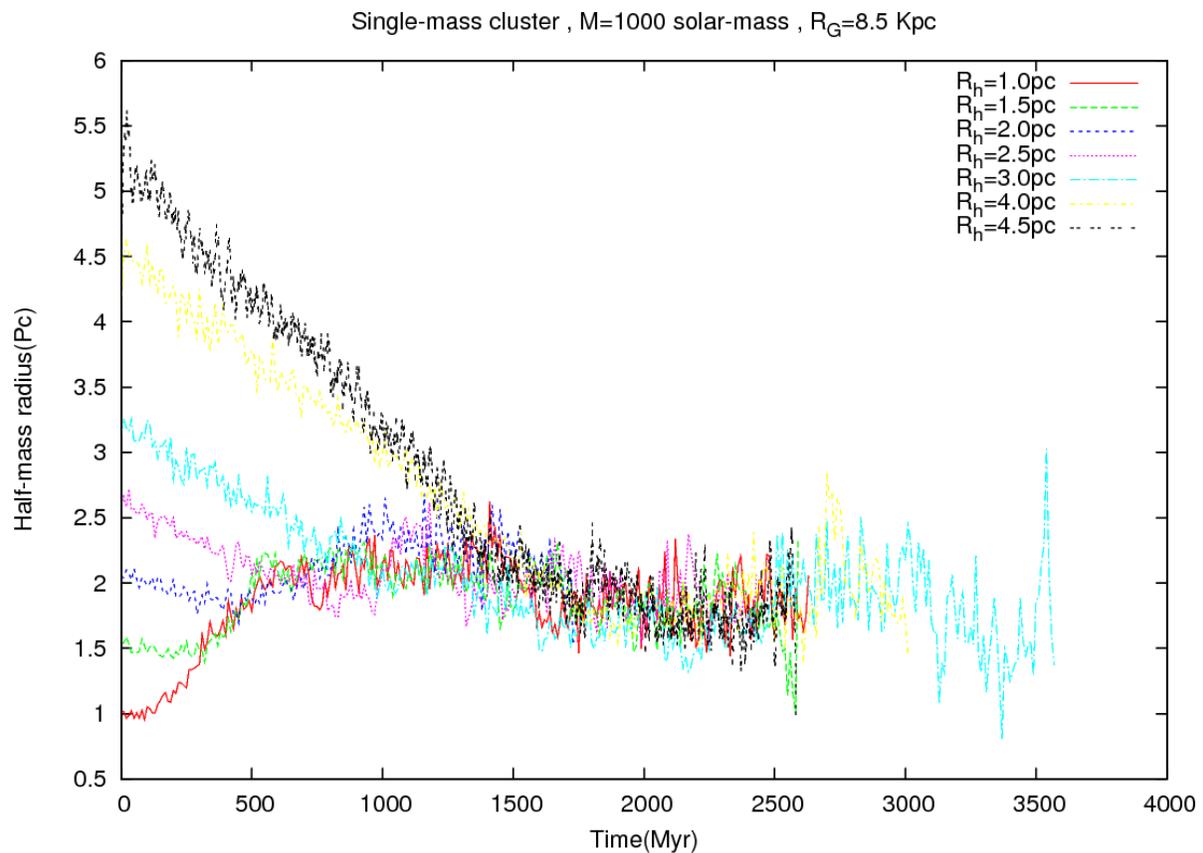

*Figure 2. 4*

These single-mass clusters have initial total mass of 1000 solar-masses and galactocentric distance of 8.5 Kpc in the Allen-Santillan tidal field. The initial half-mass radii vary from 0.5 pc to 4.5 pc.

4. **Cluster's velocity and position**

   Information about the velocity and position of the star cluster in every time-step is one of the important outputs in Nbody6. In one of the output files in Nbody6 code the position ($\vec{R}$) and the velocity ($\vec{V}$) of the star cluster is given at different times.

5. **Information about the stars**

   In Nbody6's outputs in addition to information about cluster's position and velocity, the information about every star's position and velocity in every time-steps, exists. In these files position of stars (three columns), their velocity (three columns), their mass and their kinetic and potential energy is calculated. For example a single-mass cluster with initial mass of 3000 solar-masses and initial half-mass radius of 1 pc at a galactocentric distance about 8.5 kpc in the Allen-santillan tidal field is simulated. Figure 2. 5 tells us how many stars exist in every radius at a time of 450 Myr. Figure 2. 6 gives information about the number of stars in different radii but at a time of 700 Myr which is the stationary time for this cluster.

   In single-mass clusters the column related to the stars masses is 1, because the mass of every star is 1 solar-mass in every time-step. But for multi-mass clusters the values for the stars masses are different and depend on the initial mass function (IMF). The histogram of mass in every time-step can be plotted to study the evolution of the mass function.

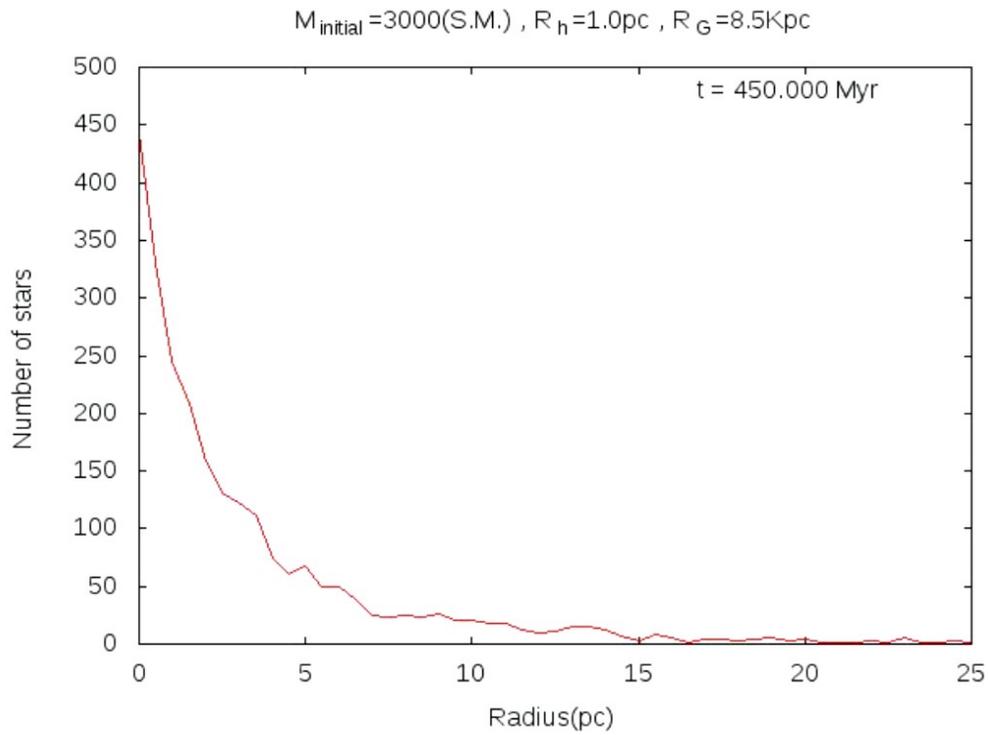

*Figure 2. 5*

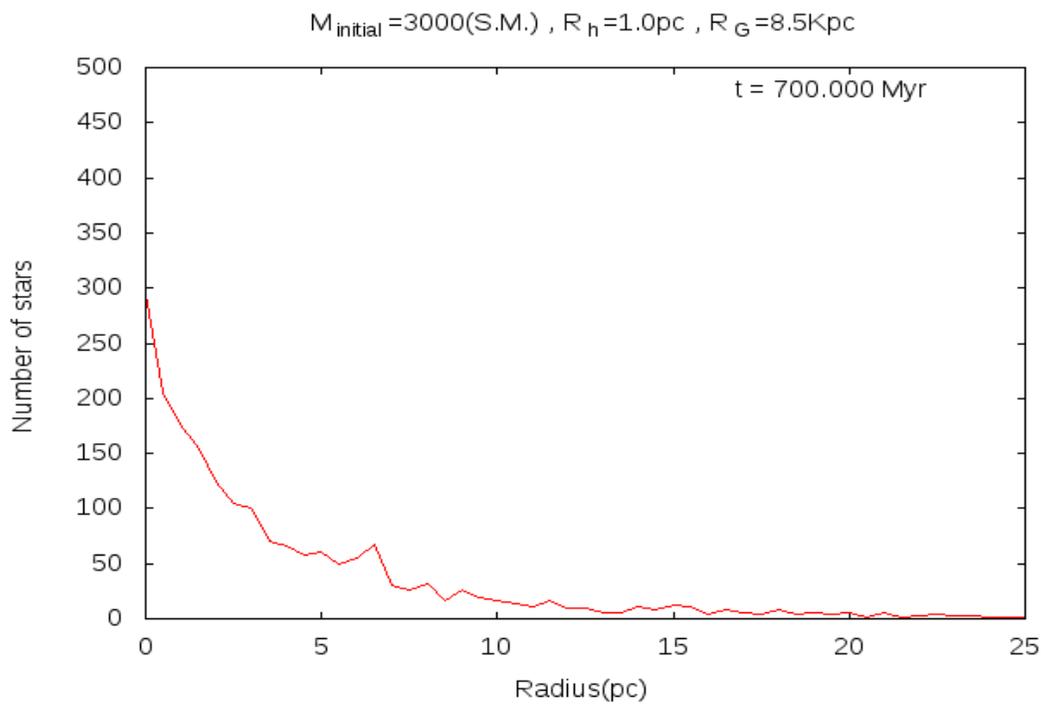

*Figure 2. 6*

# Chapter3

# Results of single-mass clusters

**1. Introduction**

In this chapter the results of single-mass clusters (mass of all stars are equal to 1 solar-mass) will be discussed.

All the simulated clusters are single-mass, with initial total mass of $M_0$ and initial half-mass radius of $r_{h0}$ at galactocentric distance of $R_G$. They move in circular orbit with initial velocity of 220 km/s round the galaxy center, in the Allen-Santillan tidal field. Generally, the magnitude and direction of initial position and velocity of simulated single-mass clusters (in Nbody6's coordinate) is:

$$\vec{R} = (8500, 0, 0)$$
$$\vec{V} = (0, 220, 0)$$

The Plummer model is used as the initial density profile.

All simulated star clusters have the same initial conditions and differ just in their initial total mass ($M_0$), initial half-mass radius ($r_{h0}$) and galactocentric distance ($R_G$). In this chapter we want to find a relation between the cluster's dissolution time and these initial parameters ( $M_0$ , $r_{h0}$ , $R_G$ ). For this aim, two parameters are kept constant and the third one varies, to see how this parameter affects the dynamical evolution and dissolution time of clusters. This method is repeated for the two other parameters, too.

$$r_{h0} = const \quad , \quad R_G = const \quad ----- \quad T_{diss}(M_0)$$

$$r_{h0} = const \quad , \quad M_0 = const \quad ----- \quad T_{diss}(R_G)$$

$$M_0 = const \quad , \quad R_G = const \quad ----- \quad T_{diss}(r_{h0})$$

The dissolution time is considered as a time when the cluster's total mass falls to 5 percent of its initial total mass:

$$T_{diss} = T_{(M=5\% \ of \ M_0)}$$

In section 2, two evolutionary time-scales important for the interpretation of our results are described. Section 3 deals with the relation between the cluster lifetime and its initial parameters. In section 4, it is shown that the cluster reaches a constant (stationary) radius. The relation between the cluster properties at the time when it reaches the stationary radius and its lifetime is explored in section 5. Section 6 summarizes our results.

**2. Evolutionary time scales**

  **2. 1 Crossing time**

  The crossing time is the time it takes for a star in a cluster traveling with the average velocity, to traverse the cluster. This will depend on the star's orbit and the size of the cluster. The crossing time may be defined as the characteristic radius of the cluster, $r_h$ , divided by the mean velocity, $v_m$ , of stars with respect to the cluster center. If the radius is in pc and the velocity in km/s , then the crossing time is:

$$T_{cross} \doteq 10^6 \ \frac{r_h}{v_m} \ yr \quad (1)$$

If one assumes that the Virial theorem holds (perhaps not a good assumption in young clusters, but safe in old ones), then:

$$v_m^2 \doteq \frac{1}{2} \ \frac{GM}{r_h} \quad (2)$$

and the crossing time can be written in terms of the cluster's size ( $r_h$ ) and mass ( $M$ ):

$$T_{cross} \doteq 2.1 \times 10^7 \ (\frac{r_h^3}{M})^{1/2} \ yr \quad (3)$$

Here again, the radius is in pc, and the mass is in solar-mass.

A typical mass for a Galactic globular cluster is $10^5$ solar-mass, and a typical half-mass radius is 3 pc (Harris 1996). This equation gives a typical crossing time of $3.5 \times 10^5 yr$, which is very much smaller than the age of these clusters.

In young LMC clusters, a typical mean velocity might be 2-3 km/s (Lupton *etal.* 1989), and radii 15-20 pc. For these radii, the crossing time is about $10^7 yr$. While uncertain, this number suggests that the ages of the young LMC clusters are comparable to their crossing times.[1]

In the clusters simulated in this project, the crossing time varies from 0.5 to 1.5 Myr which is very much smaller than their ages (about one billion years).

**2. 2 Two-body relaxation time[2]**

The relaxation time is the time it takes for a star to be perturbed by other stars. It is similar to the ratio of the velocity to the acceleration resulting from the perturbation.

If the stellar density is large (like globular star clusters), two-body gravitational relaxation is important in the system (it is comparable to the cluster's lifetime). These are called collisional stellar systems. In more dilute systems the relaxation could take longer than the age of the universe. These are called collisionless stellar systems.

---

1  R. A. W. Elson: *stellar Dynamics in Globular Clusters*
2  L. S. Sparke , J. S. Gallagher III : *Galaxies in the universe: An introduction*

If a stellar system consist of N stars of average mass m per star, the two-body relaxation time can be estimated in the following manner.[3]

We start with a star of mass $m_*$ in Figure 3. 1, moving at speed V along a path that will take it within distance b of a stationary star of mass m. The motion of $m_*$ is approximately along a straight line; the pull of m gives it a small motion, $V_\perp$ perpendicular to that path.

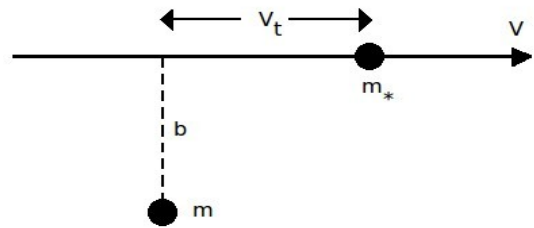

*Figure 3. 1*

If we measure time from the point of closest approach, the perpendicular force is :

$$F_\perp = \frac{Gmm_* b}{(b^2+V^2 t^2)^{3/2}} = m_* \frac{dv_\perp}{dt} \quad (4)$$

Upon integrating over time, long after the encounter, the perpendicular speed of $m_*$ is :

$$\Delta V_\perp = \frac{1}{m_*}\int_{-\infty}^{+\infty} F_\perp(t)\,dt = 2\frac{Gm}{bV} \quad (5)$$

As star $m_*$ proceeds through the cluster, many stars m will tug at it, each changing its motion by an amount $V_\perp$, but in different directions. If the forces are random, then we should add the squares of the perpendicular velocities to find the expected value of $V_\perp^2$. During time t, the number of stars m passing $m$ with separations between b and b + db is just the product of their number density n and the volume V t · 2π b db in which these encounters can take place. Multiplying by $V$ from Equation 5 and integrating over b gives the expected squared speed, after time t :

$$\langle \Delta V^2 \rangle = \int_{b_{min}}^{b_{max}} nVt \left(\frac{2Gm}{bV}\right)^2 2\pi b\,db = \frac{8\pi G^2 m^2 nt}{V} \ln\left(\frac{b_{max}}{b_{min}}\right) \quad (6)$$

we usually take $b_{min} = r_s$ and $b_{max}$ to be equal to the size of the whole stellar system. $r_s$ is a strong encounter radius. If the change in the star's potential energy is at least as great as its starting kinetic energy, then the strong encounter happens. This requires :

$$\frac{Gm^2}{r} \geq \frac{mV^2}{2}, \text{ which means } r \leq r_s \equiv \frac{2Gm}{V^2}$$

---
3  PH217: *Aug-Dec 2003*

After a time $T_r$ such that $\langle \Delta V^2 \rangle = V^2$, the star's expected speed perpendicular to its original path becomes roughly equal to its original forward speed; the 'memory' of its initial path has been lost.

Defining $\Lambda = (b_{max}/b_{min})$, then :

$$T_r = \frac{V^3}{8\pi G^2 m^2 n \, \ln(\Lambda)} = \frac{1}{8\pi} \frac{N^{1/2} R^{3/2}}{G^{1/2} m^{1/2} \ln\left(\frac{b_{max}}{b_{min}}\right)} \quad (7)$$

The number density of stars is $n \sim \frac{N}{R^3}$, R is the size of the system and the total mass of the system is $M = Nm$. Noting that :

$$V^2 = \frac{GM}{R} = \frac{GNm}{R} \quad (8)$$

A convenient version of the two-body relaxation time is the reference relaxation time. This is the two-body time calculated for the mean density inside the half-mass radius. It is defined as (Spitzer, 1987):

$$T_{rh} = 0.138 \frac{N^{1/2} r_h^{3/2}}{G^{1/2} m^{1/2} \ln(\Lambda)} \quad (9)$$

For Globular star clusters the two-body relaxation time is $\sim 10^9$ yr, so they are collisional systems. A galaxy like ours has a two-body relaxation time much larger than the age of the Universe, and is hence collisionless.

The two-body relaxation time is most relevant to the evolutionary processes taking place in old globular clusters, while the crossing time and violent relaxation time are most relevant in young globular clusters such as the $\sim 10^7$ year old rich star clusters in the LMC. The timescales are not independent. For instance, external processes such as disk shocking may accelerate internal processes such as two-body relaxation and core-collapse. In a cluster undergoing core-collapse, the two-body relaxation time will decrease dramatically in the core as its density grows quickly.[4]

The star cluster lifetime is proportional to their two-body relaxation time and the crossing time[5]:

$$T_{diss} \sim K \left(T_{rh}\right)^x \left(T_{cross}\right)^{1-x} \quad (10)$$

Also speed of the star cluster's evolution has a relation with the inverse two-body relaxation time[6]:

$$(speed\ of\ evolution) \sim \frac{X}{T_{rh}} \quad (11)$$

---
4  R. A. W. Elson: *stellar Dynamics in Globular Clusters*
5  Baumgardt, *2001, eq 16*
6  Mark Gieles, *2010*

## 3. Relation between cluster's life time and initial parameters

As mentioned in the introduction of this chapter, In this project we are trying to find a relation between a star cluster's dissolution time and some important initial parameters ( $M_0$ , $r_{h0}$ , $R_G$ ).

### 3. 1 The relation between a star cluster's lifetime and its initial total mass

In order to find the relation between the cluster dissolution time and its initial parameters, we run several groups of simulations. In each group, all the parameters except one are kept constant. For instance, Figure 3. 2 shows two groups in which the variable parameter was the initial mass of the cluster, and the groups themselves differ by the initial half mass radius. The initial total mass of the cluster varies from 1000 to 5000 solar-masses. If the dissolution time for these clusters are calculated, then the dissolution time can be plotted versus initial total mass of the cluster in logarithmic scale. In Figure 3. 2 this plot is shown. All clusters are single-mass, move in a circular orbit at a galactocentric distance of 8.5 Kpc, round the galaxy center in the Allen-Santillan tidal field.

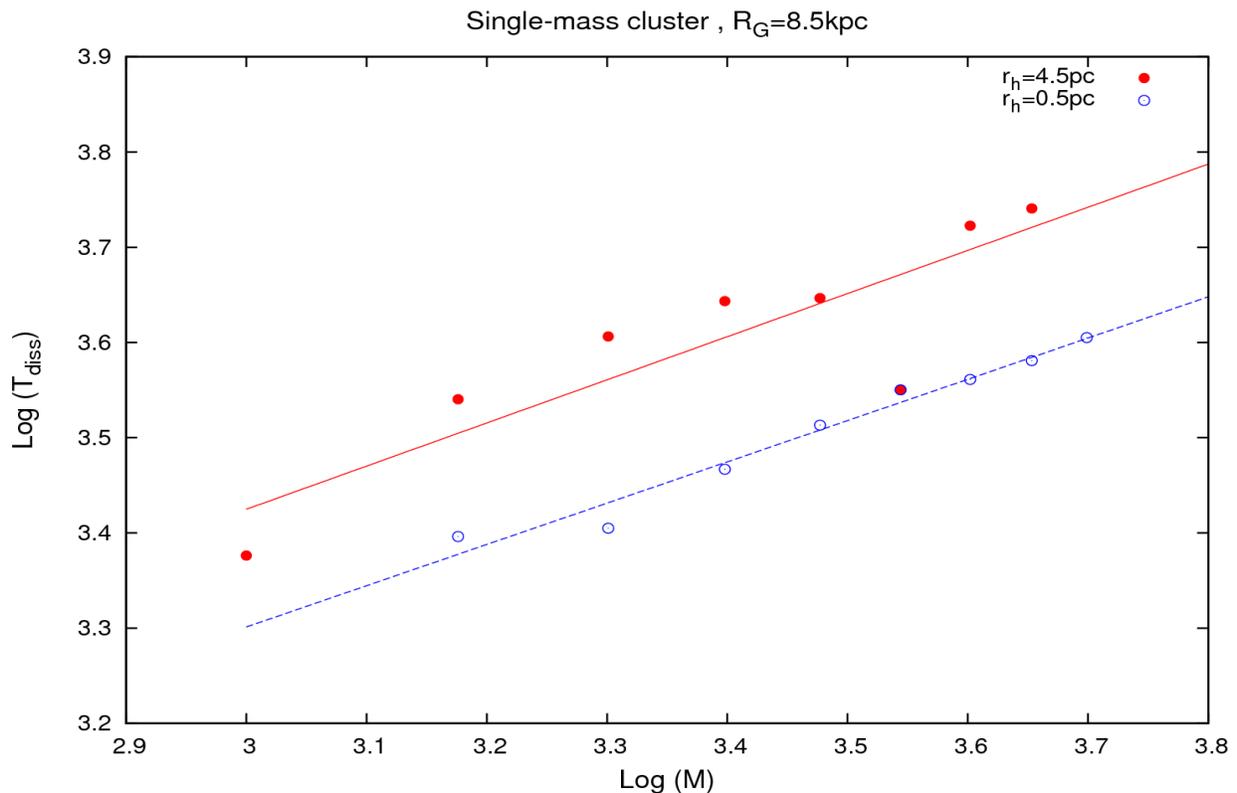

*Figure 3. 2*

For $R_G = 8.5\,kpc$ and $r_h = 0.5\,pc$ (blue point in Figure 3. 2), the fitted function is:

$$\log(T_{diss}) = (0.43 \pm 0.03)\ \log(M_0) + (2.00 \pm 0.10) \quad (12)$$

For $R_G = 8.5\,kpc$ and $r_h = 4.5\,pc$ (red point in Figure 3. 2), the fitted function is:

$$\log(T_{diss}) = (0.45 \pm 0.10)\ \log(M_0) + (2.06 \pm 0.35) \quad (13)$$

The slopes of the two function are nearly the same. For a constant half-mass radius, the dissolution time of the star clusters increases with increasing their initial total mass. That is because the two-body relaxation time increases with increasing number of stars (for single-mass clusters, N=M), so the speed of evolution decreases and the dissolution time of these clusters increases.

Also for constant initial mass, when the half-mass radius increases, the dissolution time of the clusters increases, too. That is because for constant initial mass (or constant N), the two-body relaxation time increases with increasing half-mass radius. It means that the speed of evolution decreases and the dissolution time of the clusters increases.

**3. 2 The relation between a star cluster's lifetime and its initial half-mass radius**

In this section we want to find a relation between the star cluster's dissolution time and the initial half-mass radius. For this aim the single-mass star clusters with the same initial total mass in the Allen-Santillan tidal field and at a galactocentric distance of 8.5 kpc, are simulated. These star clusters differ just in their initial half-mass radius. Figure 3. 3 shows a relation between the dissolution time of these clusters with their initial half-mass radius. The best fitted functions for these data are:

$$M_0 = 1500\ M_\odot\ ,\ R_G = 8.5\ Kpc\ :\ \log(T_{diss}) = (0.13 \pm 0.02)\ \log(r_{h0}) + (3.44 \pm 0.01) \quad (14)$$

$$M_0 = 3000\ M_\odot\ ,\ R_G = 8.5\ Kpc\ :\ \log(T_{diss}) = (0.13 \pm 0.01)\ \log(r_{h0}) + (3.56 \pm 0.01) \quad (15)$$

$$M_0 = 5000\ M_\odot\ ,\ R_G = 8.5\ Kpc\ :\ \log(T_{diss}) = (0.16 \pm 0.01)\ \log(r_{h0}) + (3.65 \pm 0.01) \quad (16)$$

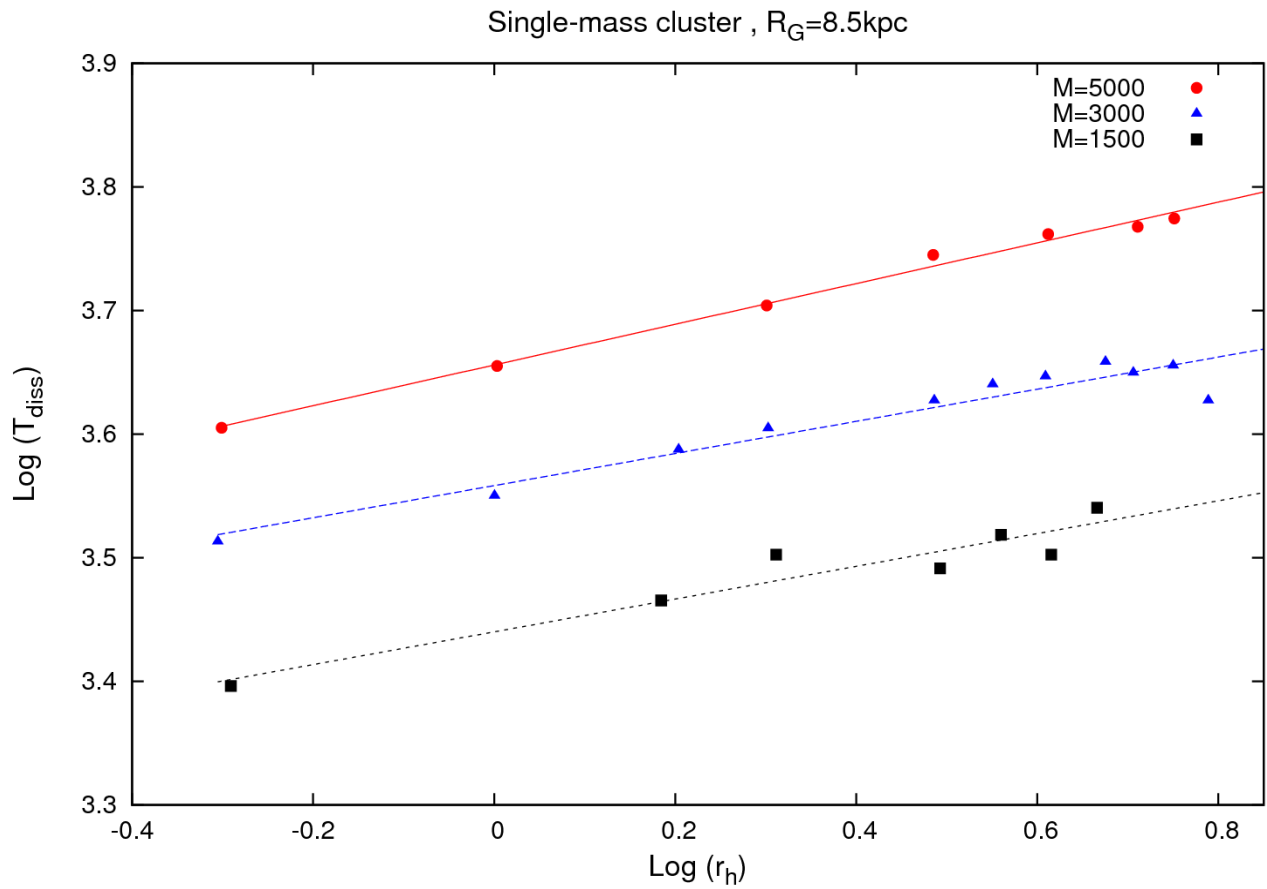

*Figure 3. 3*

It can be seen easily from Figure 3. 3 that the slopes of the different lines are nearly the same and for each specific mass, with increasing half-mass radius, the cluster's lifetime increases. From the relation that is obtained for the two-body relaxation time (eq. 7 and 9), for constant mass (or number of stars) the two-body relaxation time increases with increasing the half-mass radius. Increasing the two-body relaxation time, means that the speed of evolution decreases so the cluster's lifetime increases.

Also for fixed half-mass radius, when mass increases the cluster's lifetime increases too. That is because for fixed half-mass radius, two-body relaxation time increases with increasing the total mass (number of stars) so the speed of evolution decreases and the cluster's dissolution time increases.

**3. 3 The relation between a star cluster's lifetime and its galactocentric distance**

Single-mass clusters with the same initial total mass and half-mass radius are simulated at different distances from the Galaxy center in the Allen-Santillan tidal field. We want to find a relation between these cluster's lifetime and their distance from the galaxy center. In Figure 3. 4 this relation is shown and the best fitted function for these clusters is:

$$M_0 = 1000 \ M_\odot \ , \ r_h = 1 \, pc \ : \ \log(T_{diss}) = (0.96 \pm 0.05) \log(R_G) + (2.35 \pm 0.06) \quad (17)$$

Clusters with larger galactocentric distances, have longer lifetimes because stars in more distant clusters feel smaller tidal forces, so the evaporation rate decreases so the dissolution time increases.

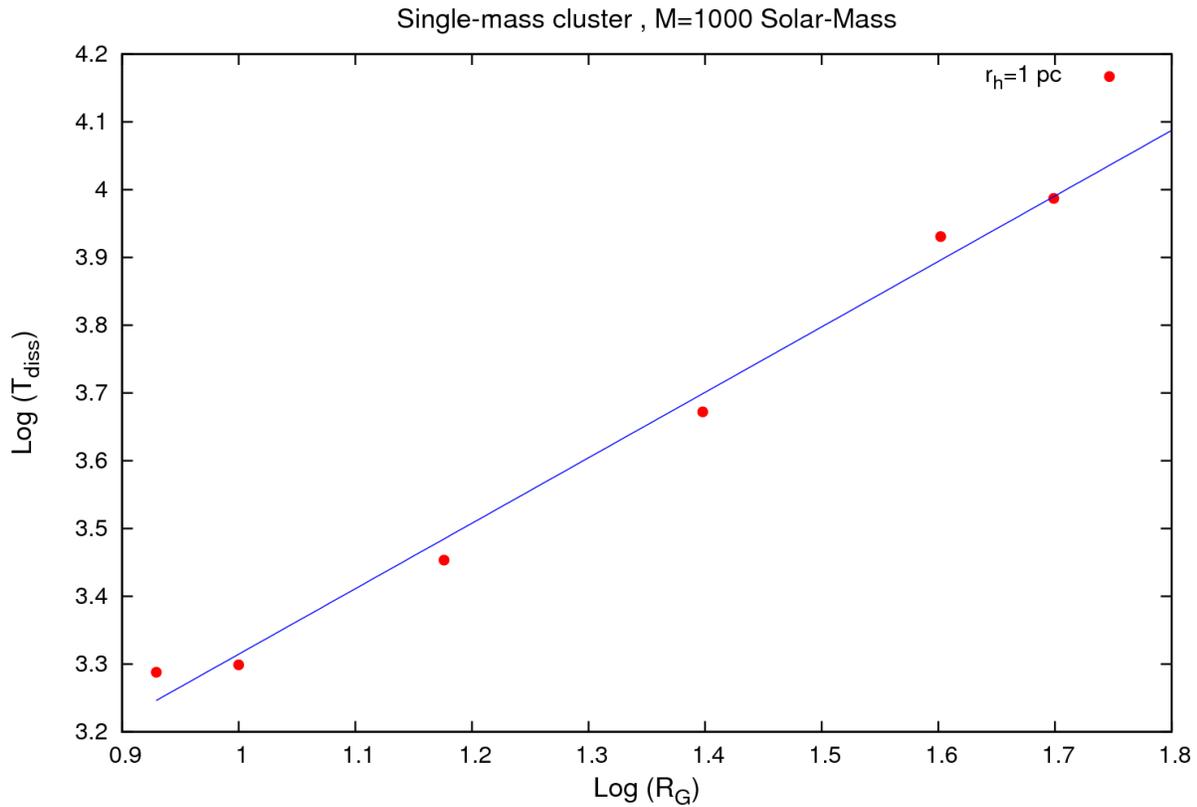

*Figure 3. 4*

213 single-mass star clusters with different initial mass and half-mass radius at different galactocentric distances in the Allen-Santillan tidal field, were simulated. For these clusters, the relation between the dissolution time and initial parameters (mass, half-mass radius and galactocentric distance) is calculated. In a table below, these equations are recorded:

| $r_{h0}=0.5\ pc$ | $R_G=8.5\ Kpc$ | $T_{diss}=100\ M_0^{0.43}\ Myr$ |
|---|---|---|
| $r_{h0}=4.5\ pc$ | $R_G=8.5\ Kpc$ | $T_{diss}=126\ M_0^{0.44}\ Myr$ |
| $M_0=1500\ M_\odot$ | $R_G=8.5\ Kpc$ | $T_{diss}=2754\ r_h^{0.13}\ Myr$ |
| $M_0=3000\ M_\odot$ | $R_G=8.5\ Kpc$ | $T_{diss}=3548\ r_h^{0.14}\ Myr$ |
| $M_0=5000\ M_\odot$ | $R_G=8.5\ Kpc$ | $T_{diss}=4467\ r_h^{0.16}\ Myr$ |
| $r_{h0}=1\ pc$ | $M_0=1000\ M_\odot$ | $T_{diss}=219\ R_G^{0.96}\ Myr$ |

Final function which is fitted to all of these clusters is :

$$\log(T_{diss})=(0.94\pm0.04)\ \log(R_G)+(0.46\pm0.03)\ \log(r_h)+(1.14\pm0.13) \quad (18)$$

Or :

$$T_{diss}=(13.80\pm1.35)\ R_G^{(0.94\pm0.04)}\ M_0^{(0.46\pm0.03)}\ Myr \quad (19)$$

In Figure 3. 5 and Figure 3. 6 the data of these clusters and the fitted function in two and three dimensions, are shown.

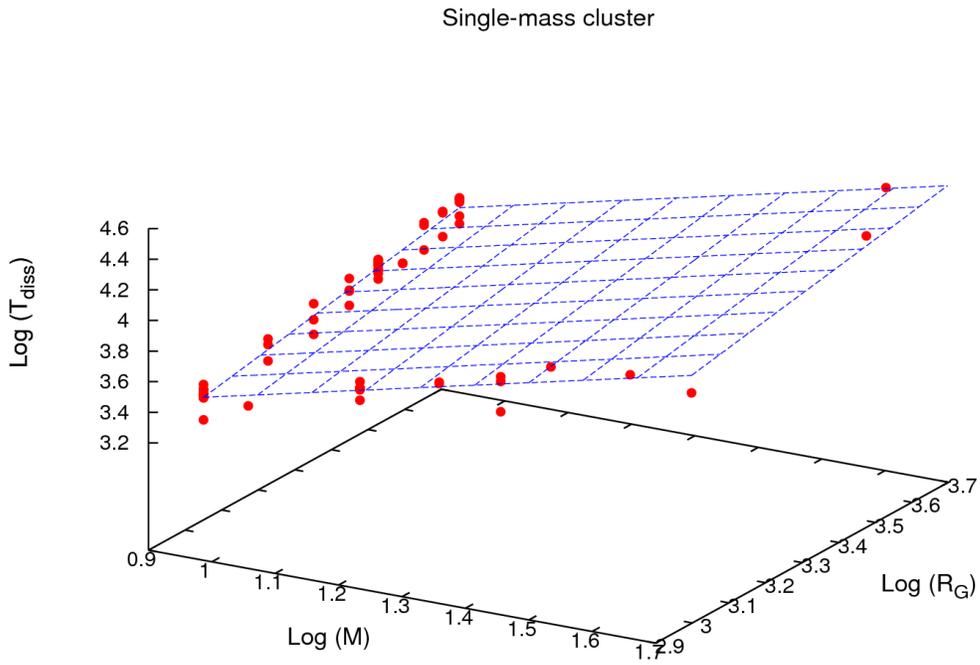

*Figure 3. 5*

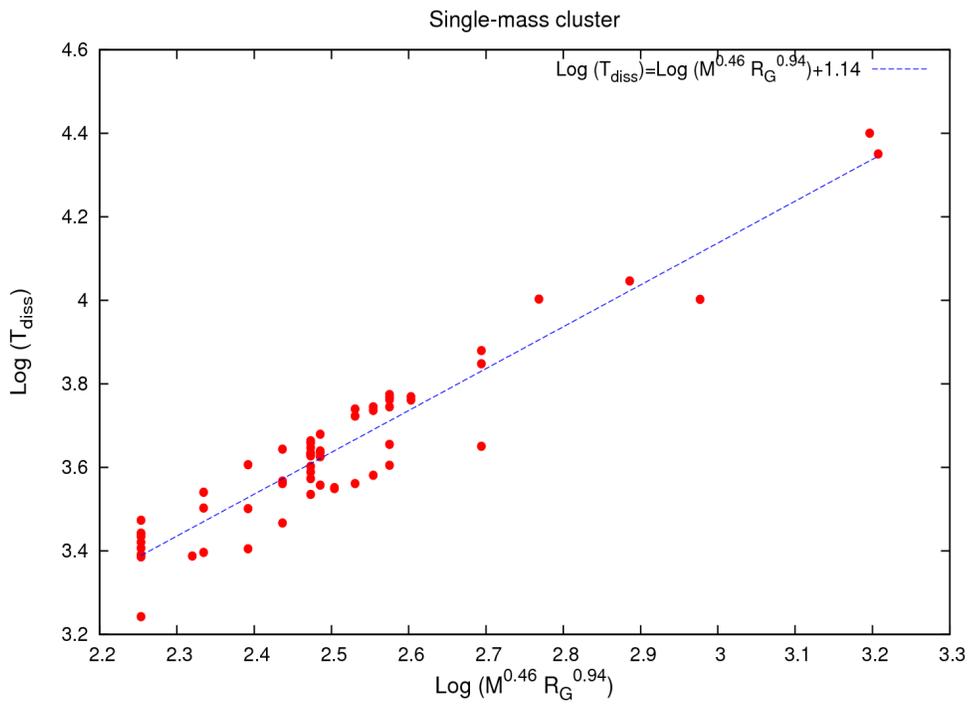

*Figure 3. 6*

**4. Stationary radius**

For the clusters with the same initial mass and galactocentric distance but different half-mass radii, the evolution of different half-mass radii shows all half-mass radii converge to the constant value. This constant value depends on the cluster's total mass and its galactocentric distance.

For example in Figure 3. 7 all clusters are single-mass with initial total mass of 1000 solar-masses. They move in a circular orbit with galactocentric distance of 8.5 Kpc round the Galaxy center in the Allen-santillan tidal field. The evolution of clusters with different initial half-mass radii is shown:

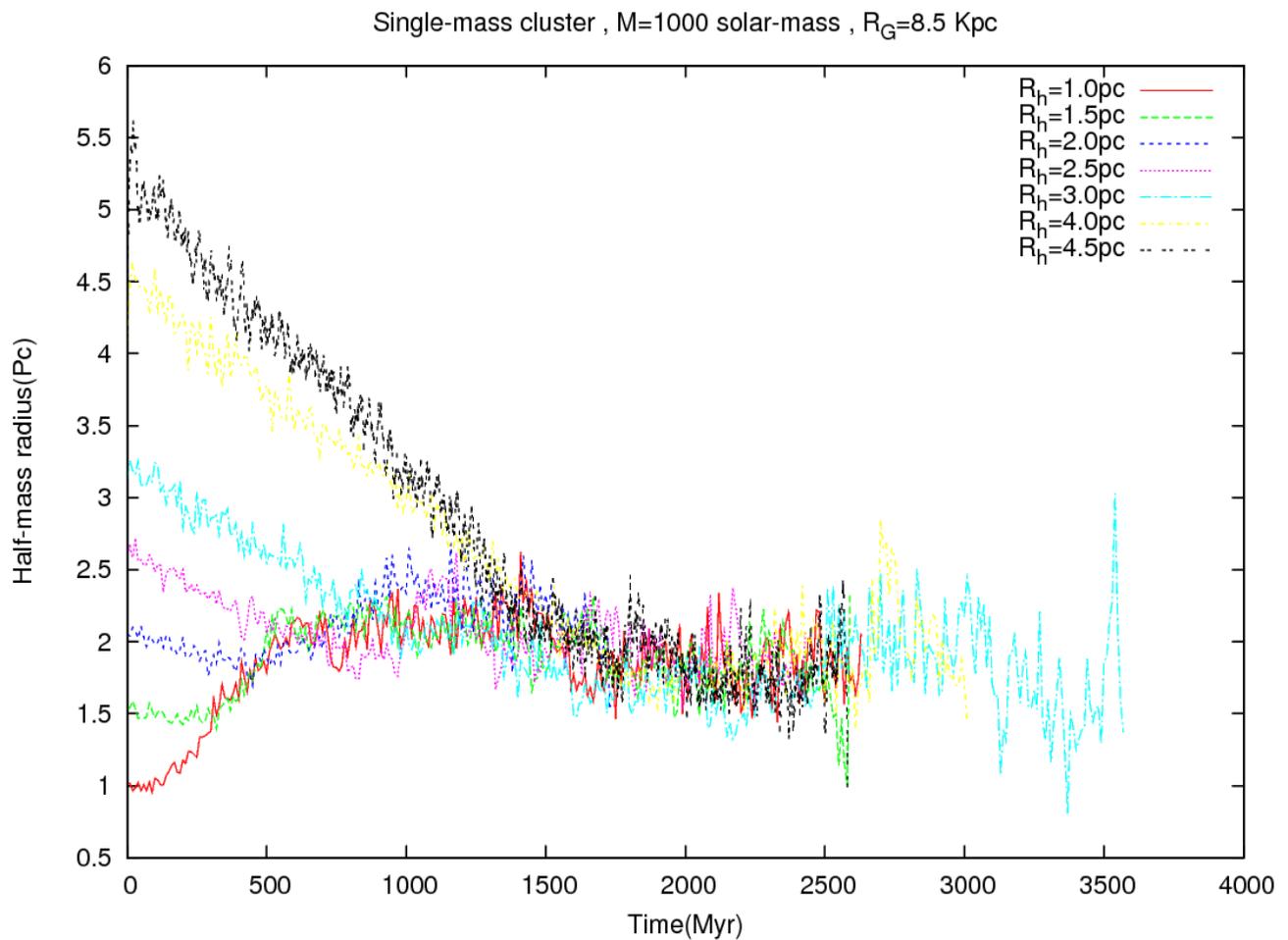

*Figure 3. 7*

It can be seen that all half-mass radii converge to a constant value of 2 pc. If the initial total mass of the cluster increases, this constant value increases too. In Figure 3. 8 all the clusters are single-mass with initial mass of 3000 solar-masses at a galactocentric distance of 8.5 kpc in the Allen-Santillan tidal field. All half-mass radii converge to a constant value of 2.5 pc.

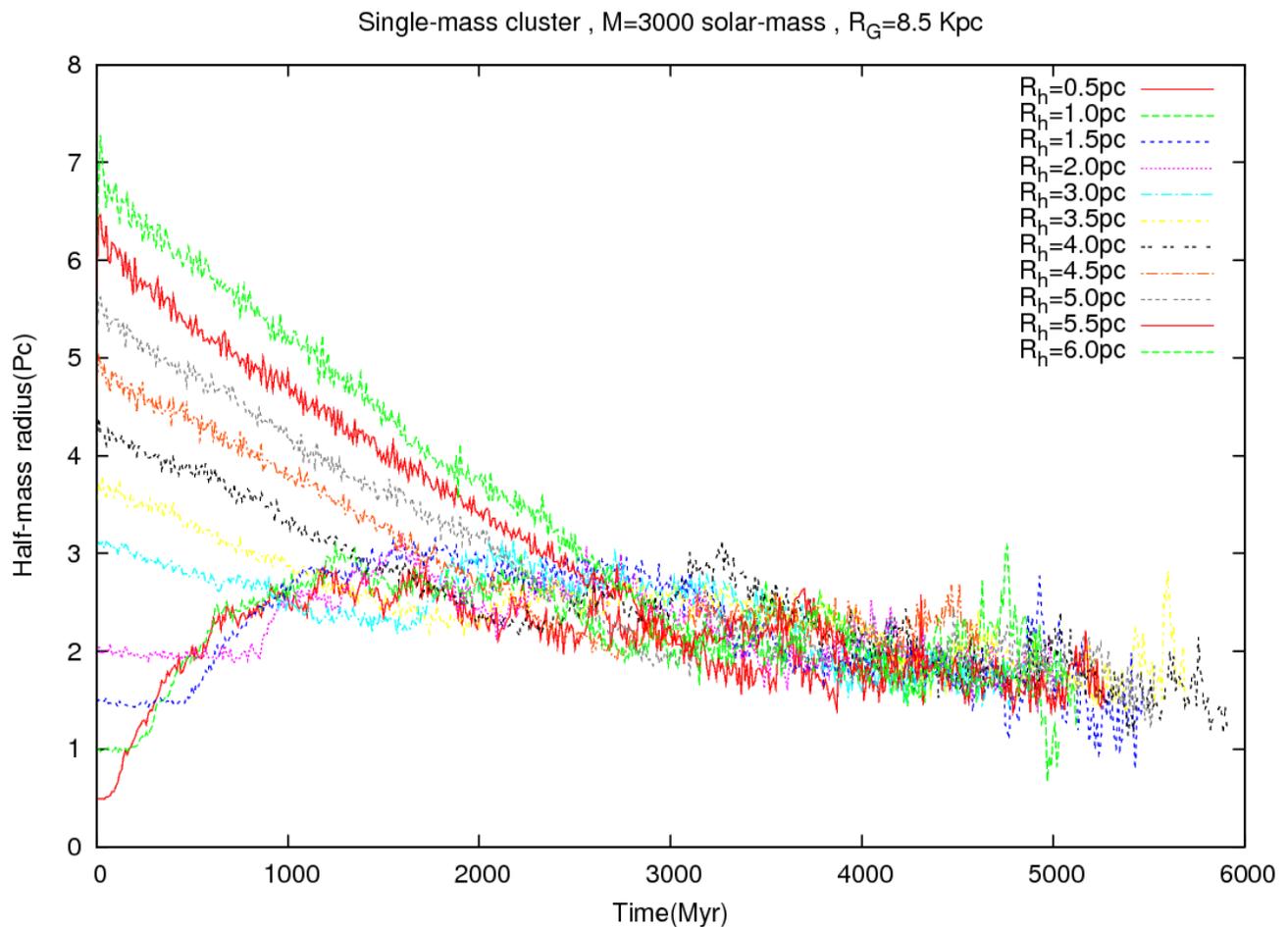

*Figure 3. 8*

And if the initial total mass of 5000 solar-masses is chosen, then this constant value reaches 3 pc. In Figure 3. 9 this feature can be seen.

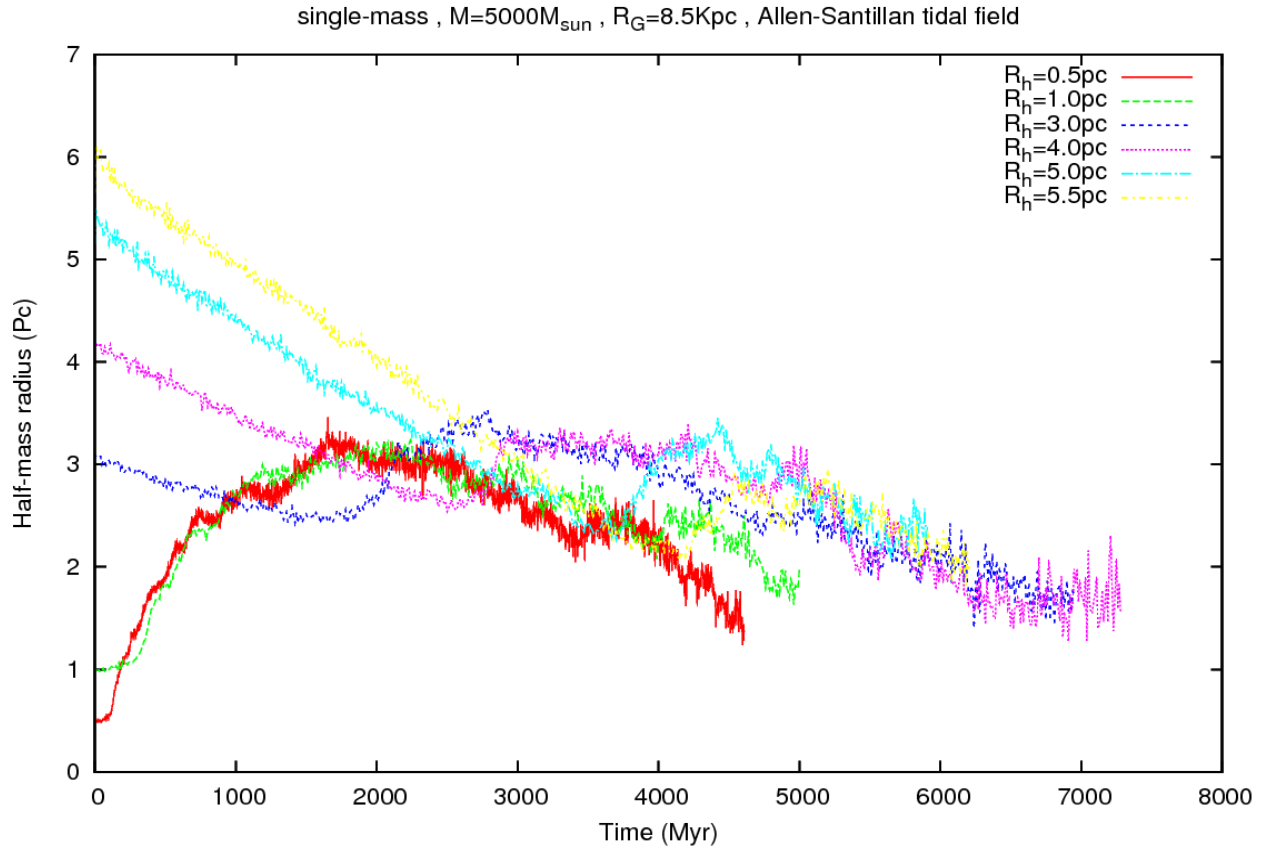

*Figure 3. 9*

The next Figures (Figure 3. 10 to Figure 3. 13) are related to the clusters with initial mass of 1000 solar-masses but at different galactocentric distance from the Galaxy center. For galactocenteric distance of 15 Kpc, 20 Kpc, 25 Kpc and 30 Kpc, this constant value is 3 pc, 3.5 pc, 4.5 pc and 5pc. Virial equilibrium implies that, reducing the half-mass radius raises the mean velocity dispersion of member stars. If two clusters of equal mass are in virial equilibrium and the first has twice the radius of the second, the second cluster must have a mean velocity dispersion half that of the first. But since these clusters have the same tidal radius, they redistribute their mass through two-body relaxation such that after core collapse the initially smaller or larger clusters do not differ in their properties, from a cluster with same initial mass.

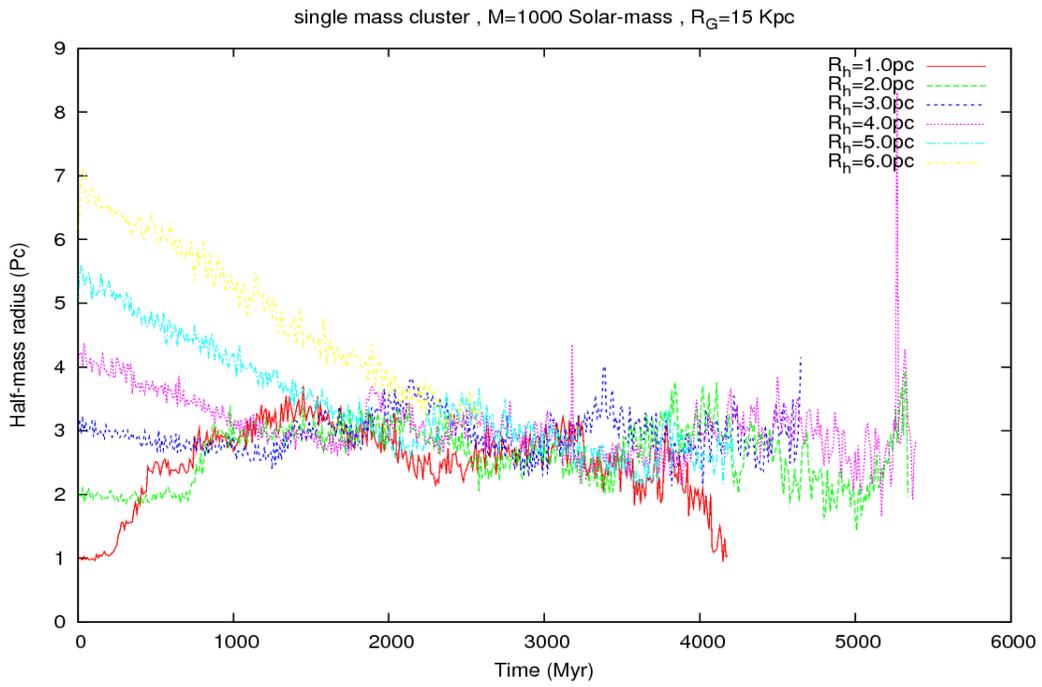

*Figure 3. 10*

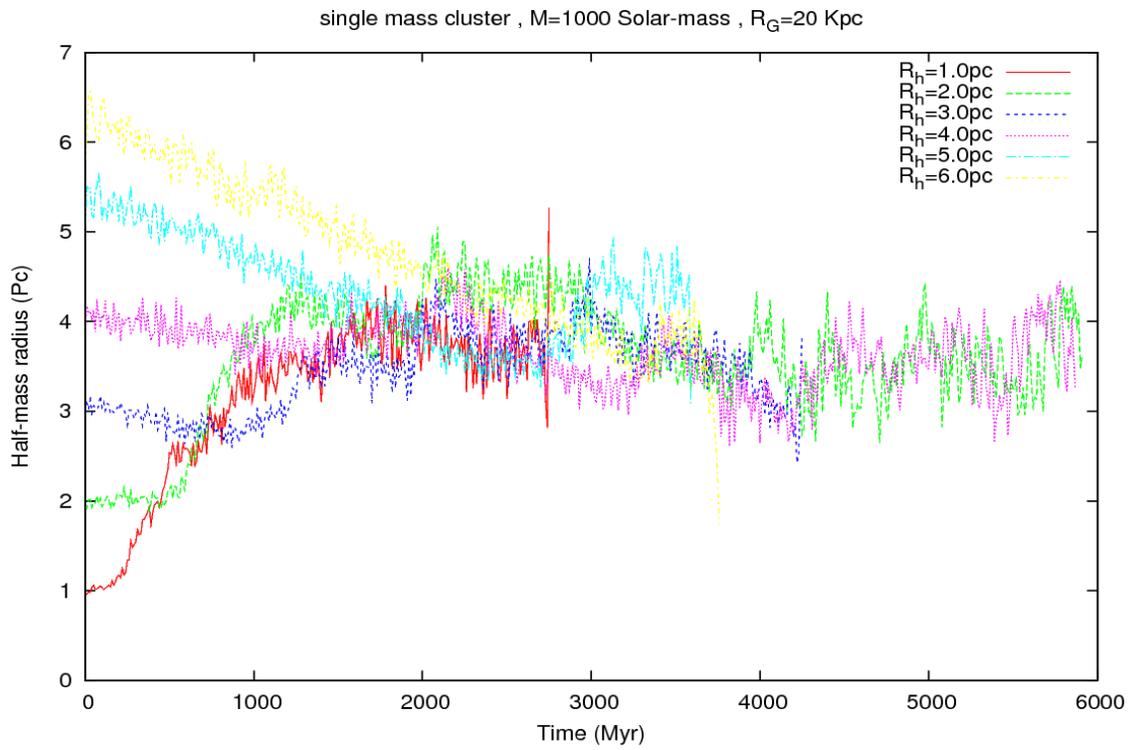

*Figure 3. 11*

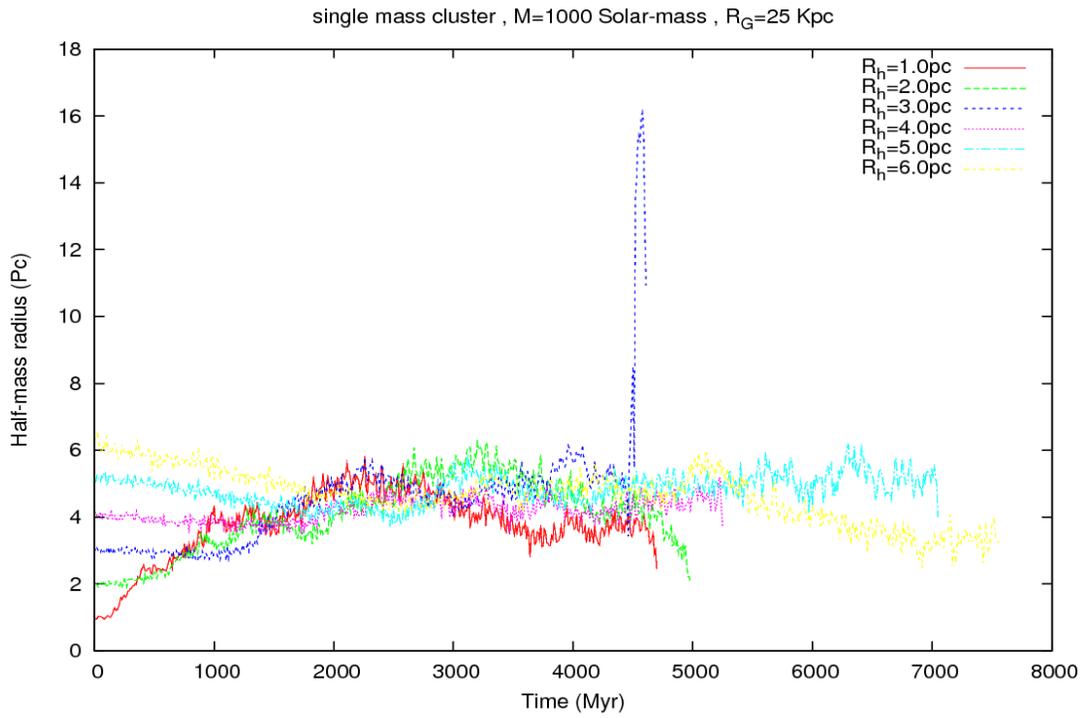

*Figure 3. 12*

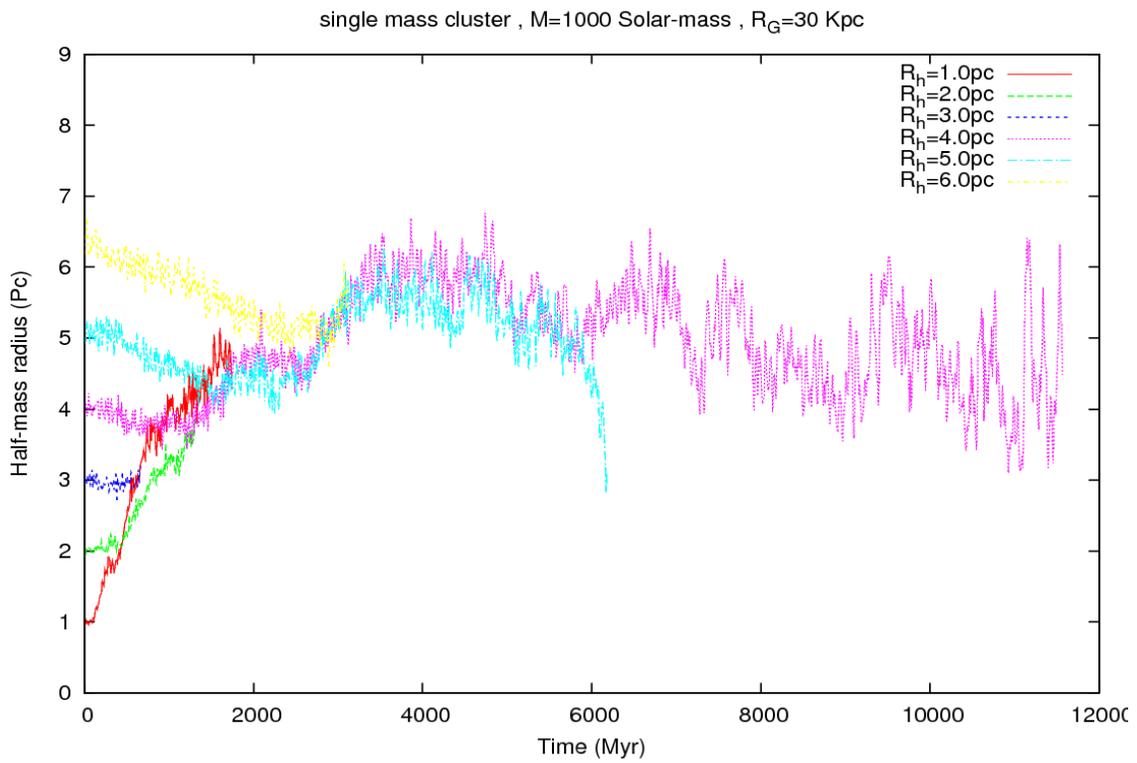

*Figure 3. 13*

All the results for the dependence of this constant value on the total mass and galactocentric distance are shown in this table. By increasing the cluster mass and Galactic distance, the asymptotic value of the half-mass radius increases. These results indicate that the evolution of half-mass radius is strongly dependent on the initial density of the cluster before core collapse, but independent of it after core-collapse. This table is the result of 55 simulation of single-mass clusters:

| $M_0\ (M_\odot)$ | $R_G\ (Kpc)$ | $R_{stat}$ |
|---|---|---|
| 1000 | 8.5 | 2 |
| 2000 | 8.5 | 2.5 |
| 3000 | 8.5 | 2.5 |
| 3500 | 8.5 | 2.75 |
| 4000 | 8.5 | 3 |
| 1000 | 10 | 2.45 |
| 1000 | 15 | 3 |
| 1000 | 20 | 3.5 |
| 1000 | 25 | 4.5 |
| 1000 | 30 | 5 |

These results show that there is no difference between the clusters with different initial half-mass radius and the cluster's stationary half-mass radius is totally independent of its initial value[7].

We name this constant value as a <u>stationary radius</u> ( $R_{stat}$ ). The time which the cluster's half-mass radius reaches to the stationary radius is called <u>stationary time</u> ( $T_{stat}$ ). If the stationary time is subtracted from the dissolution time, the stationary dissolution ( $T_{diss-stat}$ ) time is obtained. Other quantities which are calculated from these stationary parameters, are called stationary quantities. For example if the stationary half-mass radius and stationary mass are used in the calculation of the two-body relaxation time, then the stationary two-body relaxation time is obtained ( $T_{rh-stat}$ ). So for every cluster, these stationary parameters can be calculated. We want to find a general relation between the cluster's dissolution time with stationary mass and galactocentric distance.

---

7   Mark Gieles, *2010*

Recent study of star clusters shows there is a relation between the star cluster's dissolution time and its two-body relaxation time. In 2003, Baumgardt calculated this relation for multi-mass star clusters:

$$\left(T_{diss}\right) \sim C \left(T_{rh}\right)^{x} \quad (20)$$

With the old parameters ($M_0$, $r_{h0}$, $R_G$), the two-body relaxation time can be calculated. With the new (stationary) parameters ($M_{stat}$, $R_{stat}$, $R_G$), the stationary two-body relaxation time can be calculated, too. If the two-body relaxation time is plotted versus the dissolution time in logarithmic scales, no relation can be seen (Figure 3. 14).

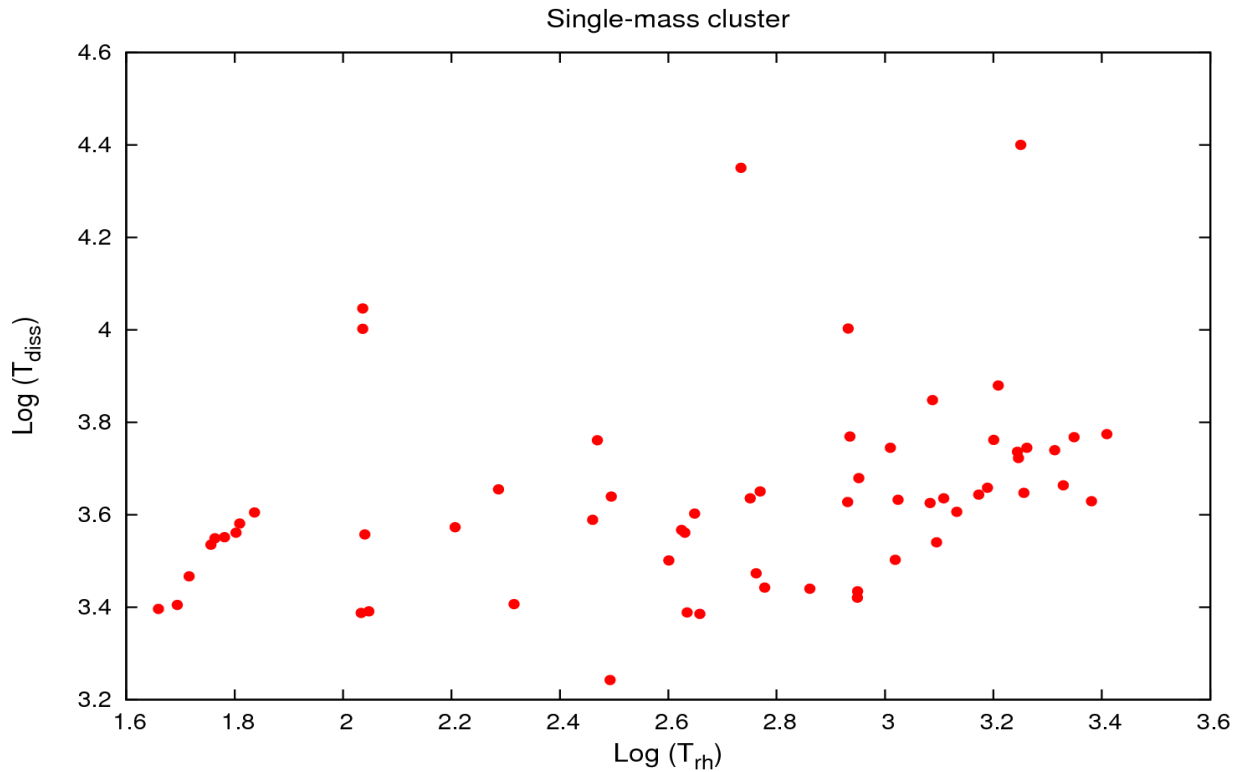

*Figure 3. 14*

Now if the stationary two-body relaxation time is plotted versus stationary dissolution time in logarithmic scales, the relation which was obtained by Baumgardt can be seen clearly (Figure 3. 15).

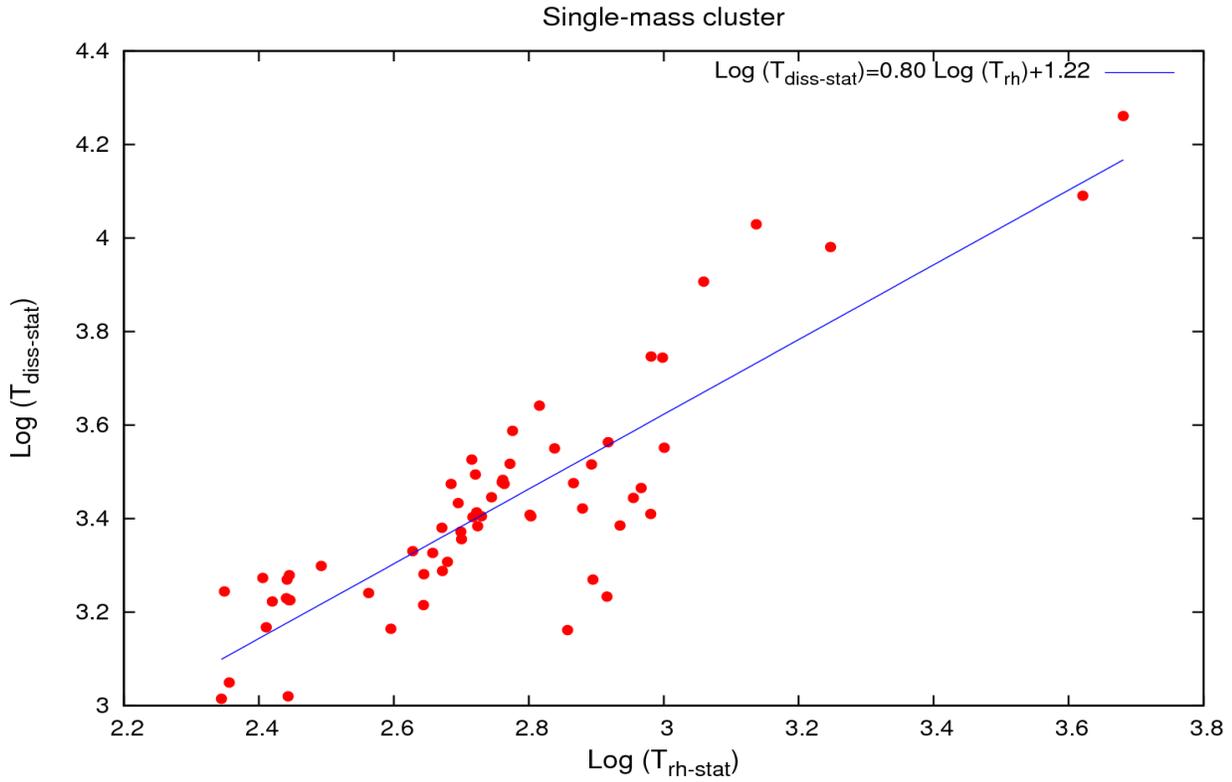

*Figure 3. 15*

The blue line is a fitted linear function on these data which is:

$$(T_{diss-stat}) = (16.59 \pm 1.48) \, (T_{rh-stat})^{(0.80 \pm 0.06)} \quad (21)$$

So with this definition of half-mass radius, mass and time (stationary parameters), we can safely study single-mass star clusters with the simplification. At the end of this chapter we will see that the results of the relation between cluster's dissolution time and initial parameters is nearly the same as the results for the relation between stationary dissolution time and stationary parameters.

**5. The relation between a star cluster's lifetime and the stationary parameters**

With this new definition, the relation between the cluster's stationary dissolution time can be obtained with cluster's stationary mass and galactocentric distance. This relation is true for all clusters with any given initial total masses and half-mass radii at different galactocentric distances from the center of the galaxy:

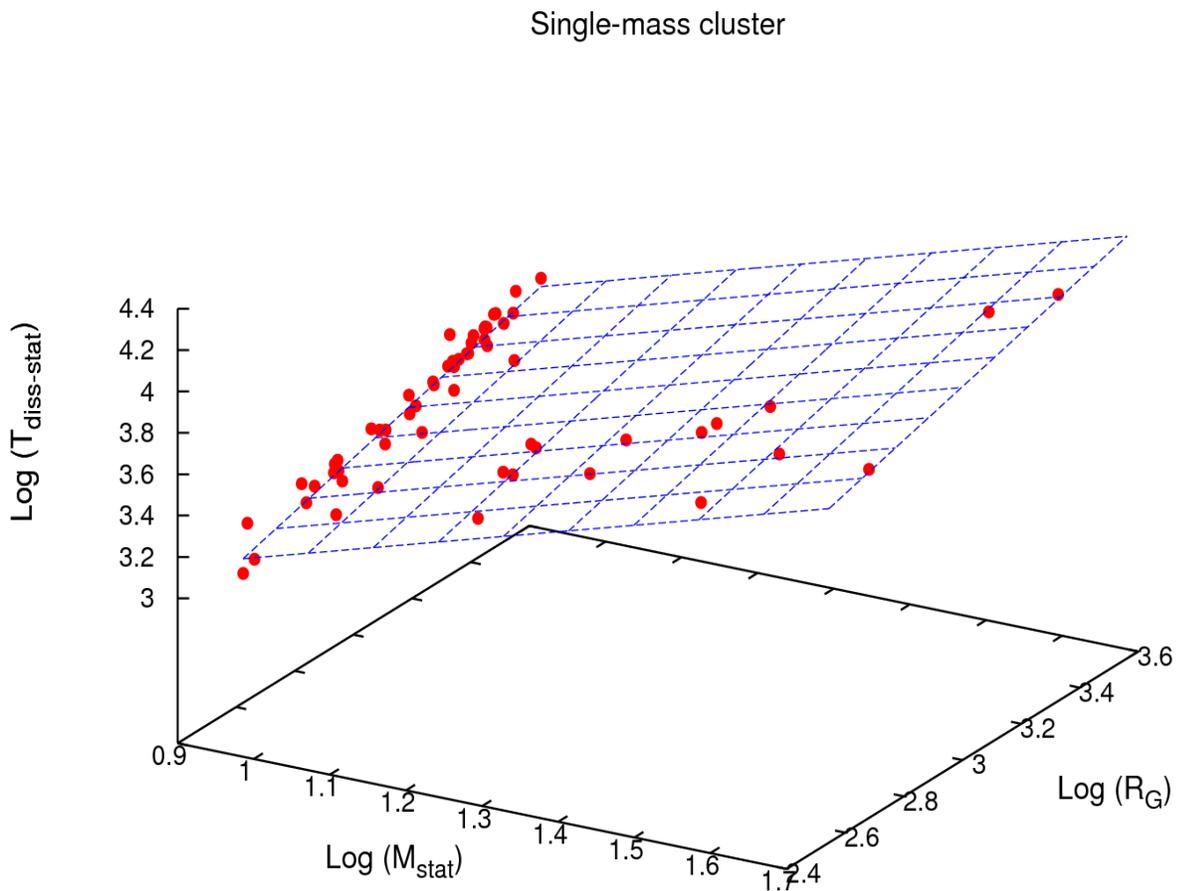

*Figure 3. 16*

and in two dimension:

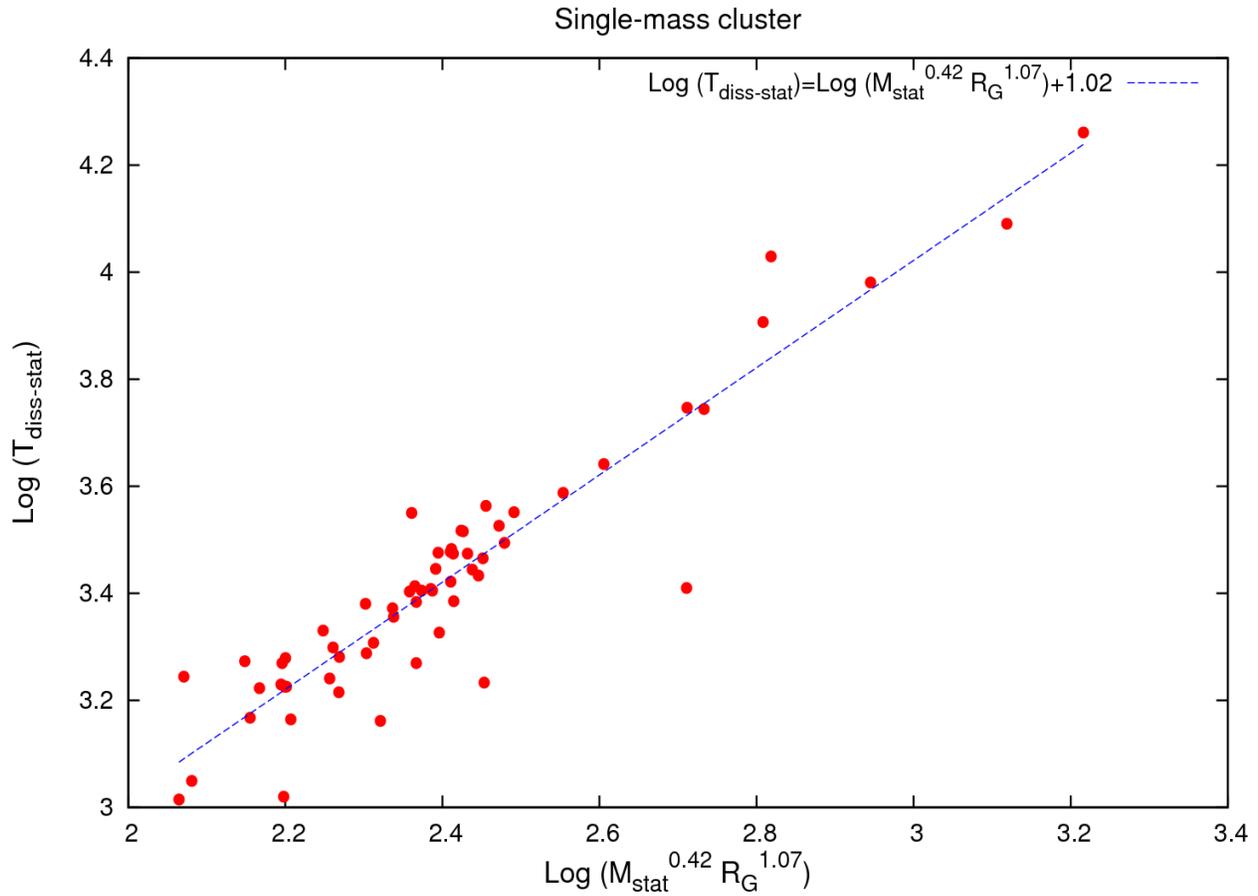

*Figure 3. 17*

The fitted plan in these data is:

$$\log(T_{diss-stat}) = (1.07 \pm 0.05) \log(R_G) + (0.42 \pm 0.04) \log(M_{stat}) + (1.02 \pm 0.16) \quad (22)$$

## 6. Conclusion

Comparing equation 19 and 23, shows that the relation between cluster's dissolution time and its initial parameters is nearly the same as the relation between cluster's stationary dissolution time and its stationary parameters.

$$T_{diss-stat} = (10.47 \pm 1.44) \; R_G^{(1.07 \pm 0.05)} \; M_{stat}^{(0.42 \pm 0.04)} Myr \qquad (23)$$

The stationary relation has a good compatibility with realistic clusters. For example, a linear relation between cluster's dissolution time and its two-body relaxation time (in logarithmic scales), exists[8]. If the stationary parameters are chosen, this relation can be seen (Figure 3. 15).

By working with stationary parameters, we will not worry about the results for dissolution time's relations, because the results of stationary parameters is similar to the results of initial parameters (eq. 19 and 23).

---

8   Baumgardt, *2001*

# Chapter 4

# Results of multi-mass clusters

**1. Introduction**

In this chapter the results of 93 simulated multi-mass clusters are discussed.

All the simulated clusters are multi-mass with Kroupa initial mass function and initial total mass of $M_0$ and initial half-mass radius of $r_{h0}$ at galactocentric distance of $R_G$. They move in circular orbits with initial velocity of 220 km/s round the galaxy center, in the Allen-Santillan tidal field. Generally, the magnitude and direction of initial position and velocity of simulated single-mass clusters (in Nbody6's coordinate) is:

$$\vec{R} = (8500, 0, 0)$$
$$\vec{V} = (0, 220, 0)$$

The Plummer model is used as the initial density profile.

## 2. The dissolution time of multi-mass star clusters

For all simulated multi-mass star clusters, their dissolution time is calculated. In figure 4. 1 the relation between the star cluster's dissolution time with their initial total mass, $M_0$, and galactocentric distance, $R_G$, is shown.

The best fit is:

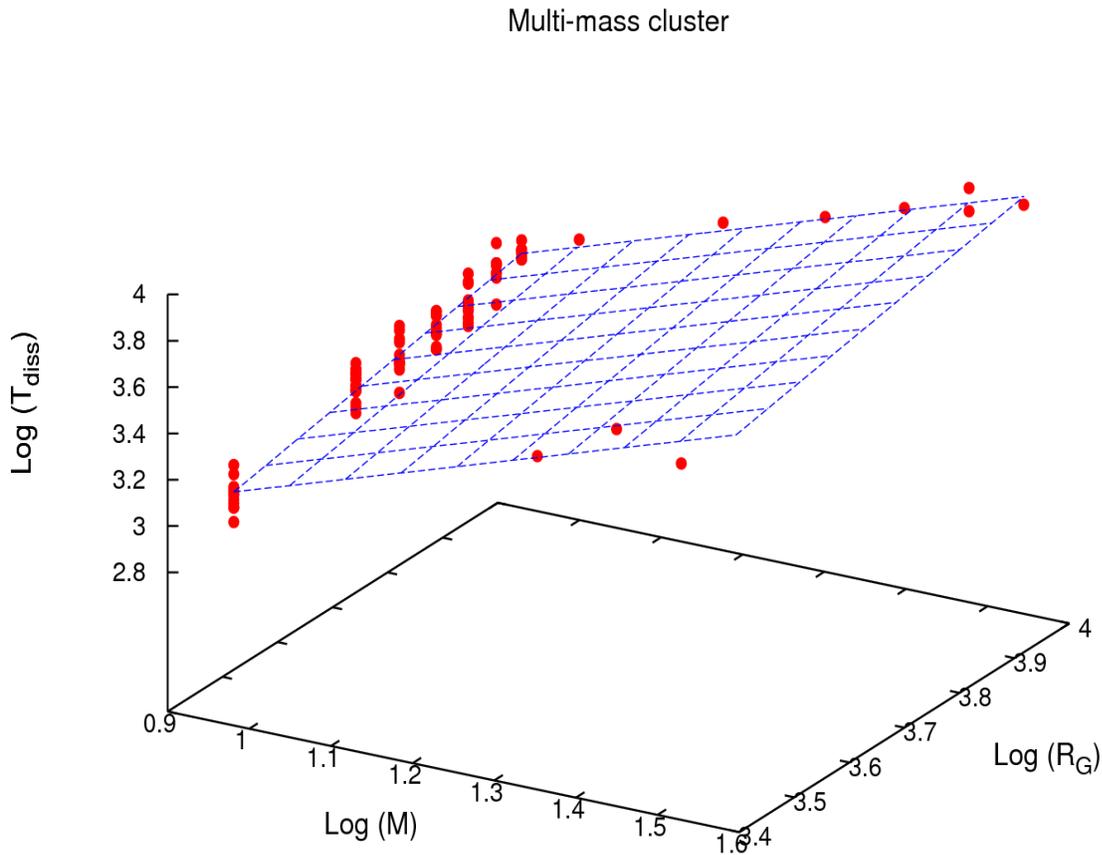

*Figure 4. 1*

$$T_{diss} = 2.34 \ R_G^{(1.14 \pm 0.05)} \ M_0^{(0.46 \pm 0.04)} Myr \quad (1)$$

This relation is very similar to the relation for single-mass clusters (chapter3, eq. 21,22).

These results show that the dissolution time increases with initial total mass and galactocentric distance. This happens because at further distances the stars of the clusters feel less tidal field and the tidal radius increases. For larger tidal radii, the rate of evaporation decreases and the cluster will have a longer dissolution time.

## 3. The mass function evolution

The initial mass function (IMF) is the histogram of stellar masses. For studying a group of stars such as star clusters, the IMF is an important tool for studying the properties and evolution of star clusters. Observations of some globular clusters show that the slope of the mass function in low-mass-stars area has a range of different values. This means that these globular clusters experienced different dynamical evolution[1]. If we assume that all the clusters have the same IMF, then the current slopes of their mass functions indicate their dynamical evolution history.

The result of dynamical evolution of star clusters and their mass loss is that their stellar mass function flatten. In this chapter, some simulations of star clusters have revealed this relation between the fraction of the initial cluster mass loss and the slope of the mass function.

The evolution of the mass functions slopes will be discussed. The simulated clusters have a Kroupa initial mass function which is :

$$N(m)dm \sim \begin{cases} m^{-1.3} dm &, \quad m_{min} < m < 0.5 \\ m^{-2.3} dm &, \quad 0.5 < m < m_{max} \end{cases} \quad (2)$$

where N(m) dm, is the number of stars with masses in the range m to m + dm .

In our simulations 0.1 and 10, is chosen for the lower and upper mass limit. Two functions can be fitted for mass function slopes (figure 4. 2) in every time step.

---

1   Piotto & Zoccali 1999

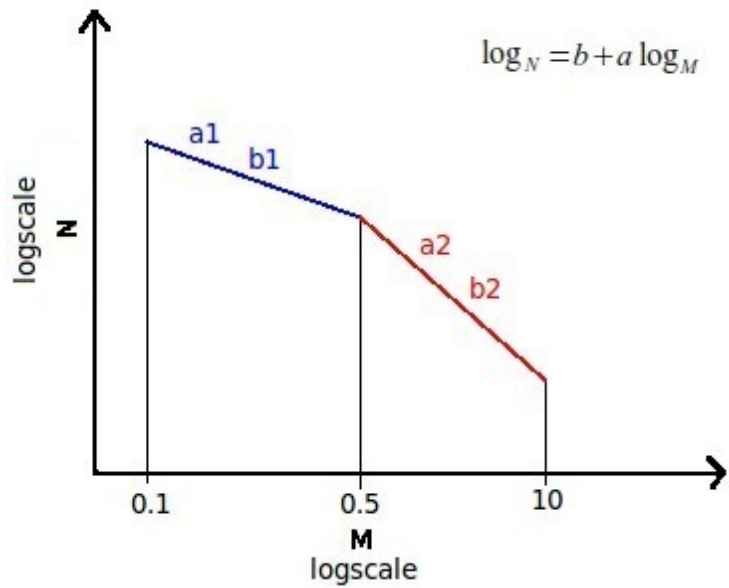

*Figure 4. 2*

a1 is the slope of the mass function for low-mass-stars (M<0.5) and a2 is the slope of the mass function for high-mass-stars (M>0.5). The evolution of these slopes will be plotted.

Consider a cluster with initial total mass of 10000 solar-masses (N=16750 stars) and half-mass radius 5 pc and galactocentric distance of 8.5 Kpc in the Allen-Santillan tidal field. In Figure 4. 3 the mass function of this cluster is plotted at 30, 60, 70 and 80 percent of the cluster's life-time. The slope of the mass function decreases during its evolution.

For this cluster, the evolution of the mass function's slopes (a1 and a2) is plotted (Figure 4. 4).

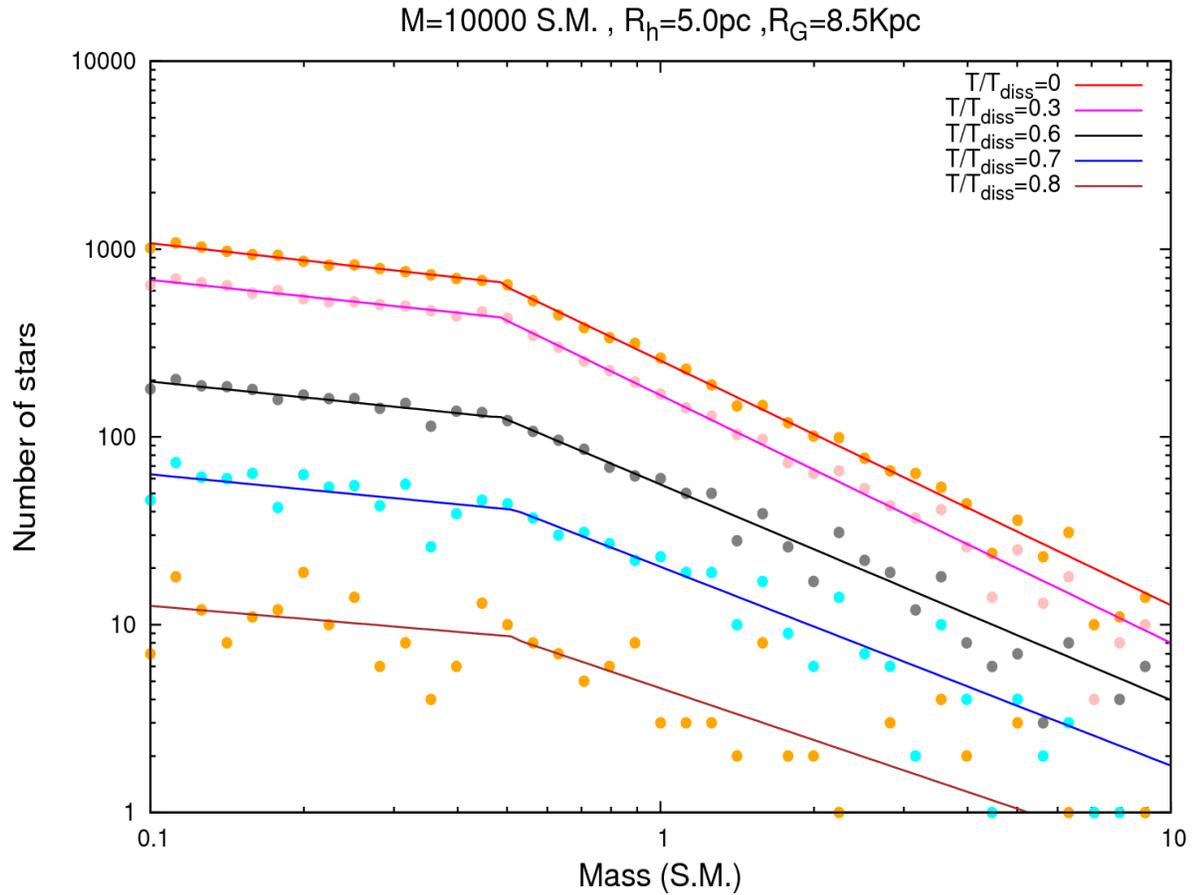

*Figure 4. 3*

It can be seen that these slopes decrease until they reach zero. This decrease in the slope magnitude shows that the number of lower mass stars which leave the cluster increases. This result satisfies the mass segregation, too. The mass segregation is a process where in gravitationally bound systems like star clusters, the high-mass-stars tend to move toward the center of the cluster while the low-mass-stars tend to move farther away from the cluster's center. During the close encounters, the energy and momentum are exchanged between the members. From the principle of equipartition of kinetic energy, there is a statistical tendency for the kinetic energy of the members to equalize during an encounter .

The kinetic energy is proportional to the mass and the square velocity. Thus the lightest stars will have higher velocity and go further from the center of the cluster and the heavier stars will have lower velocity and sink into orbits closer to the center of the cluster. After a time, called relaxation time, the kinetic energies of the stars is roughly equalize.

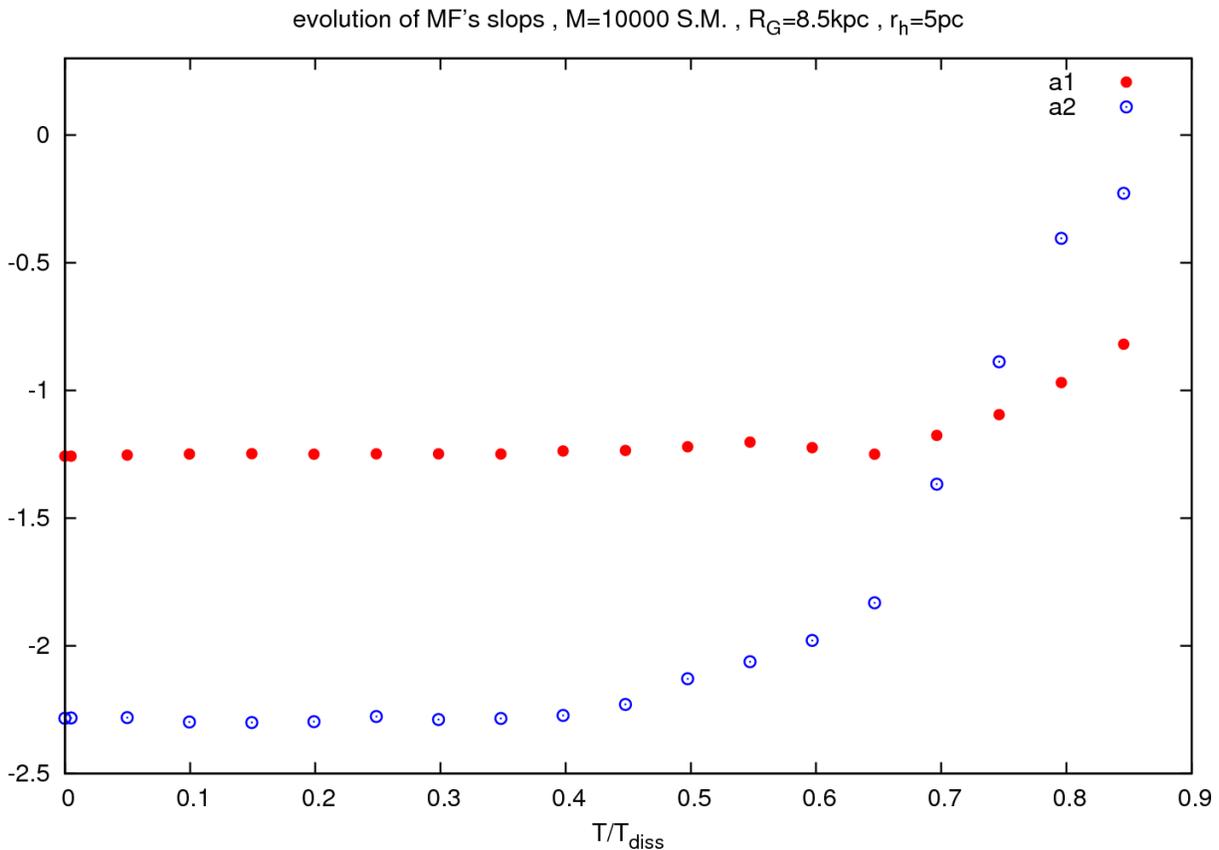

*Figure 4. 4*

In Figure 4. 5 the mass function of a cluster with initial total mass of 10000 solar-masses (N=16000), half-mass radius of 5 pc and at a galactocentric distance of 30 kpc, is plotted at 30, 60, 70 and 80 percent of the cluster's life-time.

In figure 4. 6 the evolution of the slopes of the mass function (a1 and a2) for this cluster is shown.

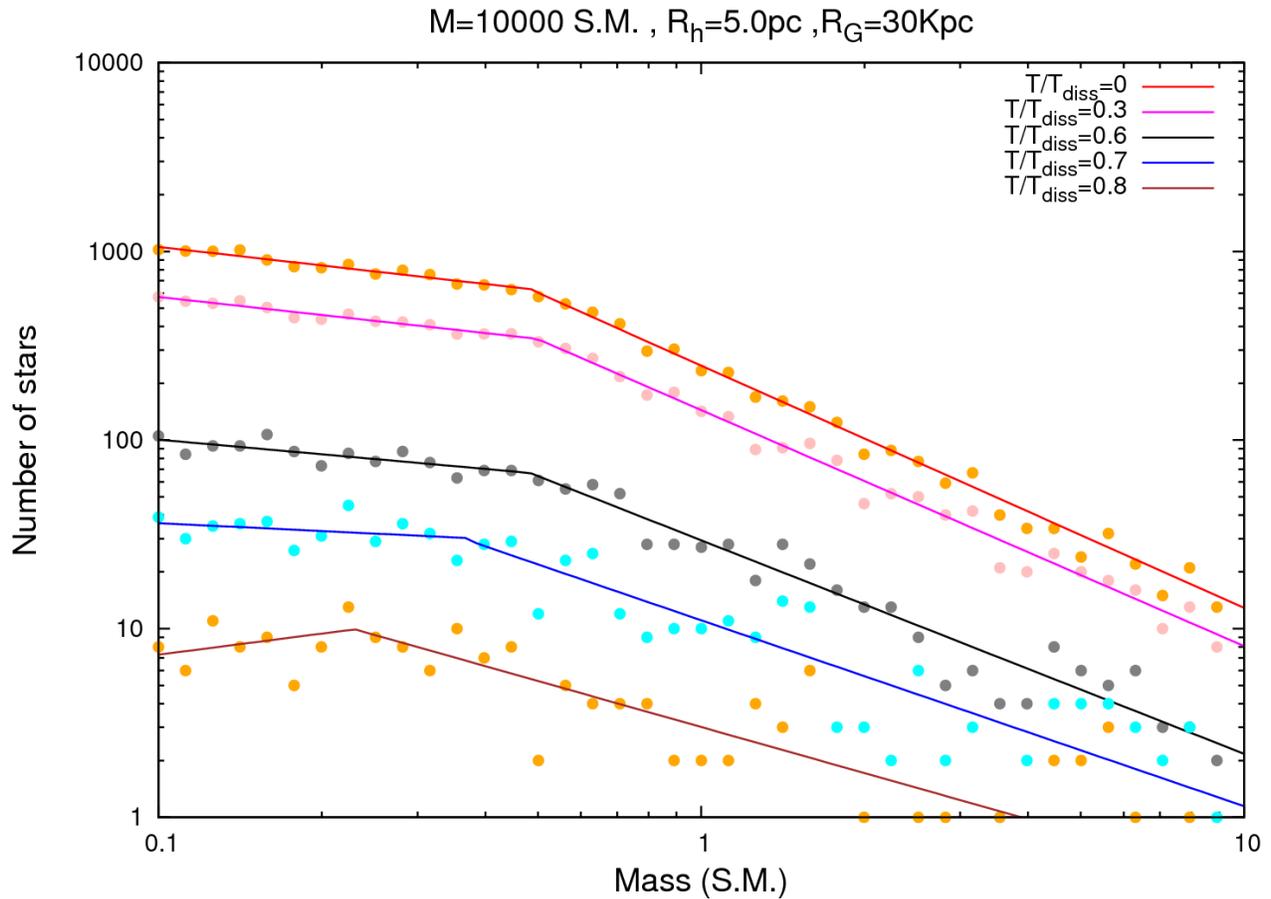

*Figure 4. 5*

For this cluster at a galactocentric distance of 30 Kpc, it can be seen that the mass at which the break in the mass function occurs decreases with time, from 0.5 solar-masses at the beginning of the simulation to 0.2 solar-masses at 80 percent of the cluster's lifetime.

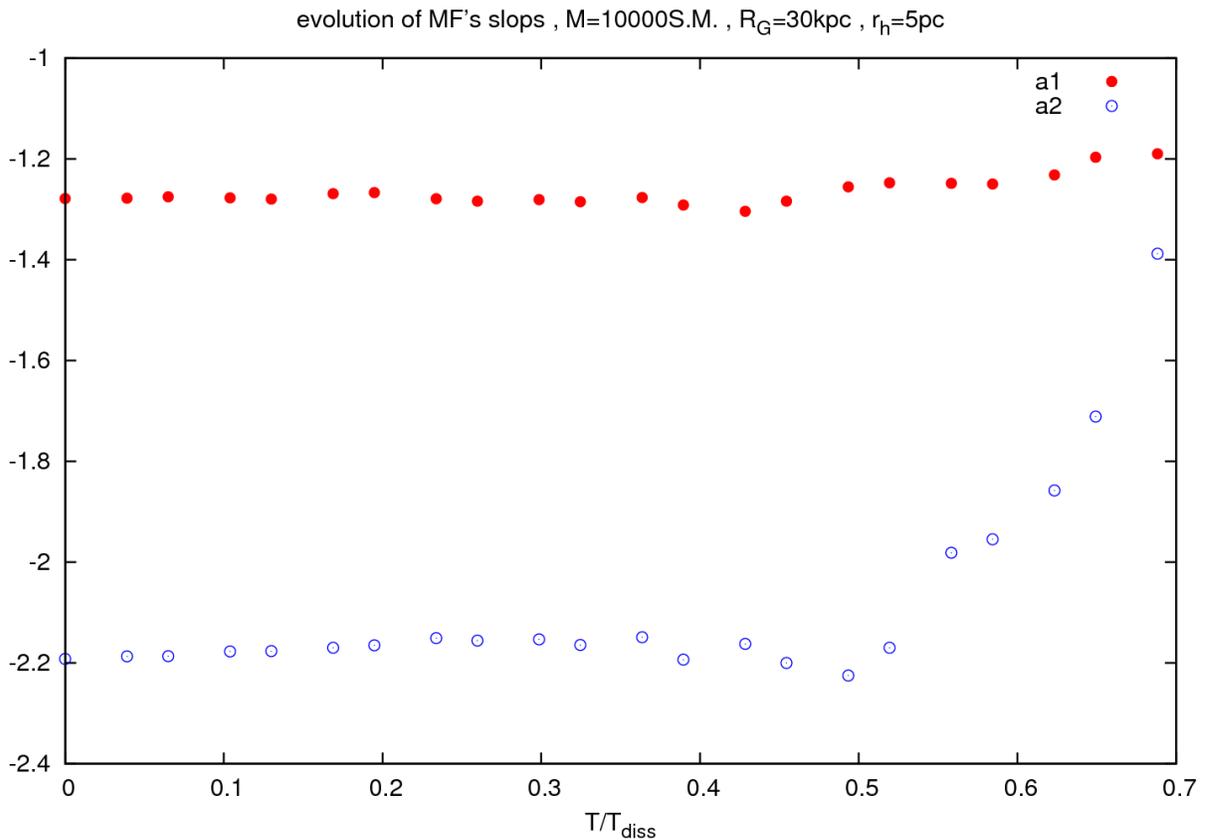

*figure 4. 6*

The next three figures (Figure 4.6 to 4.8) show the mass segregation in the cluster which is shown in Figure 4. 3 and 4. 4. These plots are generated from the stellar positions at 25, 50 and 75 percent of the cluster's lifetime. The blue points are stars with masses more than 0.5 solar-masses (high-mass-stars) and the red points are the stars with masses less than 0.5 solar-masses (low-mass-stars). It can be seen that the high-mass-stars tend to move near the cluster's center and low-mass-stars tend to move to the outer layers of the cluster.

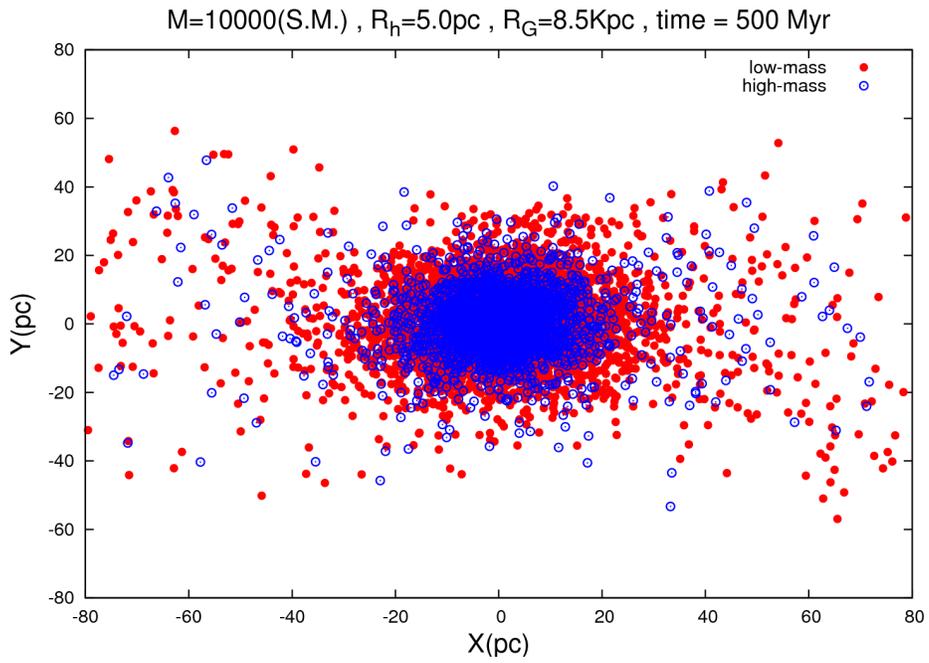

*Figure 4. 7*

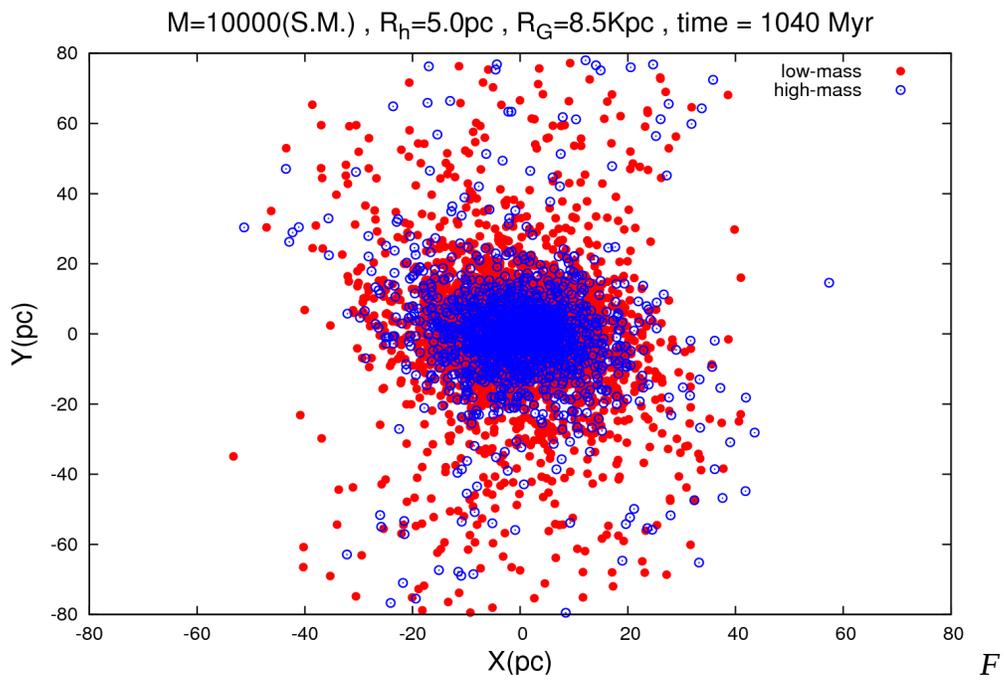

*Figure 4. 8*

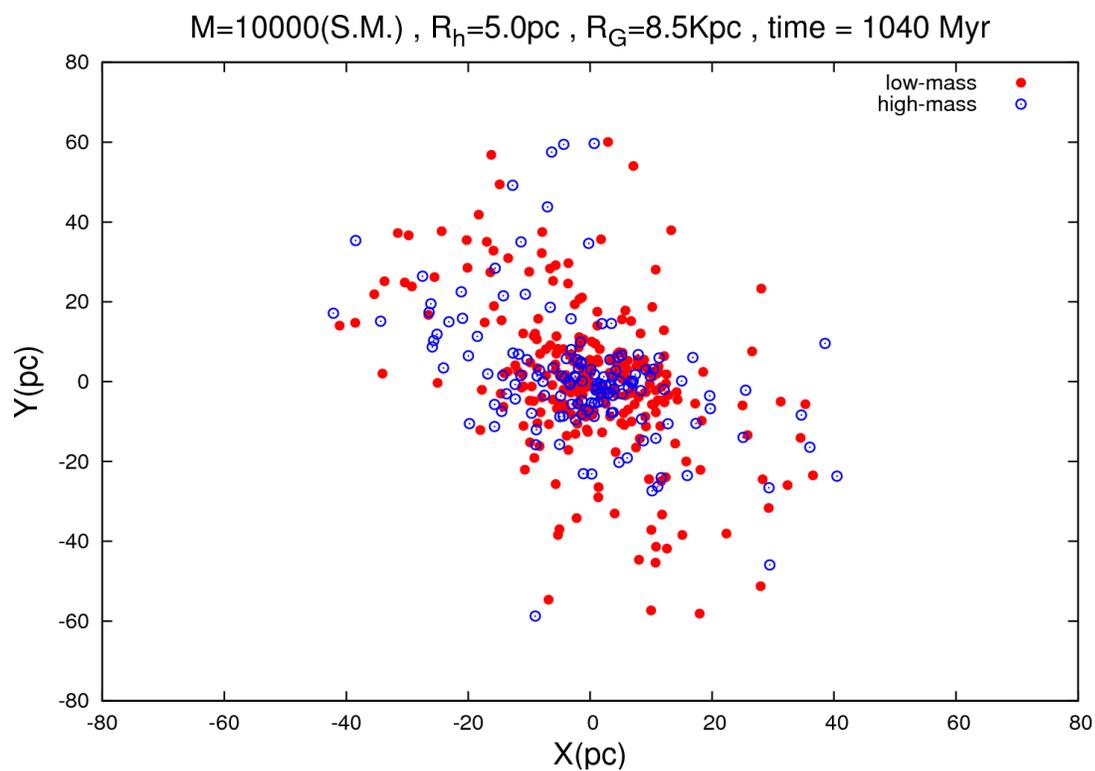

*Figure 4. 9*

## Acknowledgments

First of all I would like to thank my supervisor, Dr. Hossein Haghi for his endless advice and Akram Hasani for her useful conversations on this work.

My special thanks go to professor Yousef Sobouti, the founder of IASBS whose intelligence and vision helped many people at IASBS to study and work in an excellent environment. I have learned a lot from him during my M.Sc. Program.

Also I would like to thank professor Jan Palous, for his many helpful suggestions, and Dr. Richard Wunsch for his patience, the numerous ideas, the good advice and comments.

Thanks Jim for his many helpful suggestions and comments.

Samir, thanks for everything that you have given me and all the great times that we have shared.

Finally I would like thank my family who have always supported me in all steps of my life.

## Examination Committee

1. Dr Hosein Haghi (haghi@iasbs.ac.ir)

Supervisor of thesis, Institute for Advanced studied in basic sciences (IASBS) University

2. Prof. Yousef Sobouti (sobouti@iasbs.ac.ir)

Internal Reporter

3. Dr. Hosein Teimoorinia (teimoorinia@iasbs.ac.ir)

Internal Reporter

4. Dr. Seyed Mohammad Sadegh Movahed (m.s.movahed@ipm.ir)

External Reporter, Shahid Beheshti University

5. Prof. Babak Kaboudin (kaboudin@iasbs.ac.ir)

Representative of the University

اعضای کمیتهٔ پایان نامه

۱- دکتر حسین حقّی

(استاد راهنما - دانشگاه تحصیلات تکمیلی علوم پایه زنجان)

۲- دکتر یوسف ثبوتی

(دانشگاه تحصیلات تکمیلی علوم پایه زنجان)

۳- دکتر حسین تیموری‌نیا

(دانشگاه تحصیلات تکمیلی علوم پایه زنجان)

۴- دکتر سید صادق موحّد

(دانشگاه شهید بهشتی تهران)

۵- دکتر بابک کبودین

(نماینده دانشگاه)